\documentclass[aps,pra,twocolumn,showpacs,floatfix]{revtex4-1}

\usepackage{epsfig,amsmath,amssymb}
\usepackage{graphics} 
\usepackage{subfigure}
\usepackage{color}
\usepackage{multirow}
\usepackage{mathrsfs}
\usepackage{natbib}

\begin{document}

\title
{
A high-order study of the quantum critical behavior of a frustrated spin-$\frac{1}{2}$ antiferromagnet on a stacked honeycomb bilayer
}

\author
{R. F. Bishop$^{1,2\,}$}
\email{raymond.bishop@manchester.ac.uk}
\author
{P. H. Y. Li$^{1,2\,}$}
\email{peggyhyli@gmail.com}
\affiliation
{$^{1}$School of Physics and Astronomy, Schuster Building, The University of Manchester, Manchester, M13 9PL, United Kingdom}
\affiliation
{$^{2}$School of Physics and Astronomy, University of Minnesota, 116 Church Street SE, Minneapolis, Minnesota 55455, USA}


\begin{abstract}
We study a frustrated spin-$\frac{1}{2}$ $J_{1}$--$J_{2}$--$J_{3}$--$J_{1}^{\perp}$ Heisenberg antiferromagnet on an $AA$-stacked bilayer honeycomb lattice.  In each layer we consider nearest-neighbor (NN), next-nearest-neighbor,  and next-next-nearest-neighbor antiferromagnetic (AFM) exchange couplings $J_{1}$, $J_{2}$, and $J_{3}$, respectively.  The two layers are coupled with an AFM NN exchange coupling $J_{1}^{\perp}\equiv\delta J_{1}$.  The model is studied for arbitrary values of $\delta$ along the line $J_{3}=J_{2}\equiv\alpha J_{1}$ that includes the most highly frustrated point at $\alpha=\frac{1}{2}$, where the classical ground state is macroscopically degenerate.  The coupled cluster method is used at high orders of approximation to calculate the magnetic order parameter and the triplet spin gap.  We are thereby able to give an accurate description of the quantum phase diagram of the model in the $\alpha\delta$ plane in the window $0 \leq \alpha \leq 1$, $0 \leq \delta \leq 1$.  This includes two AFM phases with N\'{e}el and striped order, and an intermediate gapped paramagnetic phase that exhibits various forms of valence-bond crystalline order.  We obtain accurate estimations of the two phase boundaries, $\delta = \delta_{c_{i}}(\alpha)$, or equivalently, $\alpha = \alpha_{c_{i}}(\delta)$, with $i=1$ (N\'{e}el) and 2 (striped).  The two boundaries exhibit an ``avoided crossing'' behavior with both curves being reentrant.  Thus, in this $\alpha\delta$ window, N\'{e}el order exists only for values of $\delta$ in the range $\delta_{c_{1}}^{<}(\alpha) < \delta < \delta_{c_{1}}^{>}(\alpha)$, with $\delta_{c_{1}}^{<}(\alpha)=0$ for $\alpha < \alpha_{c_{1}}(0) \approx 0.46(2)$ and $\delta_{c_{1}}^{<}(\alpha) > 0$ for $\alpha_{c_{1}}(0) < \alpha < \alpha_{1}^{>} \approx 0.49(1)$, and striped order similarly exists only for values of $\delta$ in the range $\delta_{c_{2}}^{<}(\alpha) < \delta < \delta_{c_{2}}^{>}(\alpha)$, with $\delta_{c_{2}}^{<}(\alpha)=0$ for $\alpha > \alpha_{c_{2}}(0) \approx 0.600(5)$ and $\delta_{c_{2}}^{<}(\alpha) >0$ for $\alpha_{c_{2}}(0) > \alpha > \alpha_{2}^{<} \approx 0.56(1)$.
\end{abstract}


\maketitle

\section{INTRODUCTION}
\label{introd_sec}
Quantum spin-lattice models, in which various types of interactions
between pairs of spins compete to form different types of order in the
system, provide a rich arena in which to study a wide variety of
quantum phase transitions (QPTs)
\cite{Sachdev:2011_QPT,Sachdev:2008_QPT}.  Such competition can occur
either with or without magnetic frustration between the interaction
bonds.  In the latter case the bonds are typically spatially
anisotropic.  Simple examples of such systems comprise models
containing nearest-neighbor (NN) exchange interactions between spins
$\mathbf{s}_{i}$ on lattice sites $i$ of the Heisenberg type,
$J_{ij}\,\mathbf{s}_{i}\cdot\mathbf{s}_{j}$, all with bond strength
$J_{ij}>0$, and thus all acting to promote antiferromagnetic (AFM)
long-range order (LRO), but where the strengths $J_{ij}$ are not all
equal.  The so-called coupled-dimer magnets comprise a particularly
simple, yet important and nontrivial, class of this type.

Dimerized quantum Heisenberg antiferromagnets (HAFMs) are obtained by
placing quantum spins, with spin quantum number $s$, on a regular
$d$-dimensional lattice with an even number of spins per unit cell.
Each unit cell is divided into non-overlapping pairs of spins
(dimers).  In the limit where the intradimer exchange constant
$J_{ij}$ are much stronger than all of the corresponding interdimer
constants, the zero-temperature ($T=0$) ground-state (GS) phase of the
system is a simple paramagnetic product of non-magnetic spin singlets,
which preserves the full spin-rotational invariance.  This state has a
nonzero energy gap to the lowest-lying spin triplet excitation, formed
by breaking one of the spin-singlet dimers.  As the interdimer
exchange interactions are turned on these triplet excitations acquire
mobility on the lattice, resulting in the appearance of a spin-1
bosonic quasiparticle, viz., the triplon \cite{Schmidt:2003_spin1}.  In principle, of course,
such bosons may undergo Bose-Einstein condensation (BEC) under
suitable conditions and become superfluid.  Indeed, such a superfluid
state has been observed experimentally in the magnetic insulator
TlCuCl$_{3}$
\cite{Nikuni:2000_coupled_dimer_magnet,Cavadini:2001_coupled_dimer_magnet,Ruegg:2003_coupled_dimer_magnet,Oosawa:2003_coupled_dimer_magnet},
which is a physical realization of such a coupled-dimer magnet, in
which pairs of Cu$^{2+}$ ions are antiferromagnetically coupled to
form a crystalline network of dimers in a specific pattern.  The BEC
is induced by placing the compound in an external magnetic field,
which thereby Zeeman splits the otherwise degenerate three magnetic
triplet sub-states.  At some critical field strength the lowest-lying
triplet state intersects the GS dimer singlet, and BEC into this
triplon sub-state occurs, with the consequent appearance of the
magnetic LRO corresponding to the off-diagonal LRO that characterizes
BEC in the boson-mapped equivalent system.  The whole area of BEC in
magnetic insulators \cite{Giamarchi:2008_BoseEinstenCond} has become
one of considerable activity in recent years, at both the theoretical
and experimental levels.  The applied magnetic field here thus acts
as a chemical potential that promotes dimer spin-singlets (leaving a
hole) to spin-triplets (creating a triplon).

In principle, another way to induce magnetic LRO in a coupled-dimer
magnet, without the application of an external magnetic field, is to
increase the relative strength of the interdimer couplings $J_{ij}$
with respect to their intradimer counterparts.  For example, for all
two-dimensional (2D) bipartite lattices and with all couplings
$J_{ij}$ between NN pairs only, when {\it all} $J_{ij}$ are equal the
system will have N\'{e}el AFM magnetic LRO.  Thus, if we consider the
class of so-called $J$--$J'$ models on bipartite lattices, in
which the intradimer NN bonds all have the same strength $J'>0$
and the interdimer NN bonds all have the same strength $J > 0$, there
will clearly be same critical value of the relative strength
parameter, $(J'/J)_{c}>1$, that marks a QPT between a N\'{e}el-ordered
AFM GS phase and a dimerized paramagnetic GS phase.  Experimentally,
both in principle and sometimes in practice, such QPTs can be
driven by the application of pressure to the system.

On the 2D square lattice with NN interactions only, $J$--$J'$
models with specific arrangements of the $J'$ bonds that have been
studied include the columnar-dimer, staggered-dimer, and
herringbone-dimer models (and see, e.g., Ref.\
\cite{Fritz:2011_dimerized_AFM}).  The first two, each with two sites
per unit cell, both have the dimer $J'$ bonds parallel (say, along the
row direction of the square lattice).  Whereas in the columnar
arrangement each basic square plaquette contains either two or no
dimer $J'$ bonds, in the staggered arrangement each basic square
plaquette contains a single $J'$ dimer bond.  Finally, the
herringbone-dimer model contains four sites per unit cell with two
isolated (non-touching) dimer $J'$ bonds perpendicular to one another.
Each basic square plaquette also contains a single $J'$ dimer bond.
Interestingly, in the limit $J'/J \rightarrow 0$, both the
staggered-dimer and herringbone-dimer models become equivalent to the
HAFM on a hexagonal lattice.  Thus, both of these models interpolate
between HAFMs on the hexagonal and square 2D lattices for values of
$J'/J$ in the range $0 \leq J'/J \leq 1$.

So far we have considered coupled-dimer magnets that include
competition between bonds without frustration.  The inclusion of extra
bonds can now lead us into the realm of magnetic frustration, which
adds further to the complexity and inherent interest of these models.
For example, increasing frustration can have the effect of enhancing
the repulsive interactions between triplons.  In turn this can then
eventually lead to the stablization of incompressible phases that
break the translational symmetry of the lattice.  If such GS phases of
the system are placed in an external magnetic field, the itinerant
triplons become localized in a crystalline phase.

Such phases have been observed experimentally in the spin-gap material
SrCu$_{2}$(BO$_{3}$)$_{2}$
\cite{Kageyama:1999_bipart,Kodama:2002_Srcubo,Miyahara:2003_Srcubo},
which is well modeled \cite{Miyahara:1999_Srcubo} by the 2D
Shastry-Sutherland model \cite{Sriram:1981_Srcubo}.  This is a
spin-$\frac{1}{2}$ coupled-dimer model on a square lattice, with four
sites per unit cell, in which all NN pairs have an AFM Heisenberg bond
of equal strength $J$, and the equivalent AFM dimer bonds, all of
equal strength $J'$, join non-overlapping diagonal
next-nearest-neighbor (NNN) pairs in an orthogonal pattern.  The unit
cell thus contains two orthogonal dimers arranged across NNN diagonal
pairs.  Clearly, in the limit $J \rightarrow 0$ the model reduces to a
Hamiltonian of decoupled dimers.  This dimerized state then remains
the {\it exact} GS phase \cite{Sriram:1981_Srcubo} for all values of
$(J'/J)$ above a certain critical value $(J'/J)_{c}$.  In the opposite
limit, when $J' \rightarrow 0$, the model reduces to the pure
spin-$\frac{1}{2}$ HAFM (i.e., with NN interactions only) on the
square lattice, which has N\'{e}el AFM magnetic LRO.  In between, when
$J'=J$, the model reduces to the pure spin-$\frac{1}{2}$ HAFM on
another of the 11 2D Archimedean lattices, the GS phase of which is
also known to have N\'{e}el order
\cite{DJJFarnell:2014_archimedeanLatt}, which implies $(J'/J)_{c}>1$.
Various theoretical studies (see, e.g., Refs.\
\cite{Koya:2000_Srcubo,Corboz:2013_Srcubo}) yield
$(J'/J)_{c}\approx 1.48$.  In the Shastry-Sutherland material
SrCu$_{2}$(BO$_{3}$)$_{2}$, for which $(J'/J) \approx 1.6$, the
triplon crystalline phases show up as a series of magnetization
plateaux at unconventional filling fractions
\cite{Kodama:2002_Srcubo,Miyahara:2003_Srcubo} that are stabilized by
complex many-body interactions among the triplons
\cite{Fritz:2011_dimerized_AFM,Miyahara:2003_Srcubo,Abendschein:2008_Srcubo,Dorier:2008_Srcubo}.
Indeed, the magnetization plateaux in SrCu$_{2}$(BO$_{3}$)$_{2}$ at
low magnetization are now quite well understood in terms of triplon
bound states \cite{Corboz:2014_Srcubo,Foltin:2014_Srcubo}.

Since both BEC and crystalline phases of triplons can occur in
frustrated coupled dimer magnets subjected to an external magnetic
field, it is natural to wonder whether such systems might also exhibit
the magnetic equivalent of {\it supersolidity}, in which a stable GS
phase exhibits simultaneously both the diagonal LRO typical of a
solid and the off-diagonal LRO typical of a superfluid.  Such
field-induced spin supersolidity has particularly been investigated
for various frustrated spin-$\frac{1}{2}$ models on a square-lattice
bilayer \cite{Giamarchi:2008_BoseEinstenCond,Laflorencie:2007_SqLatt_bilayer,Schmidt:2008_SqLatt_bilayer,Picon:2008_SqLatt_bilayer,Chen:2010_SqLatt_bilayer,Albuquerque:2011_SqLatt_bilayer,Murakami:2013_SqLatt_bilayer}.

Such bilayer models provide other examples of coupled-dimer magnets.
The simplest such bilayer models comprise two layers stacked directly
on top of one another and with only NN bonds, where the intralayer
bonds all have equal strength $J_{1}$ and the interlayer (dimer)
bonds all have equal strength $J_{1}^{\perp}$.  Such models on the
square lattice, where the bonds compete without frustration, have been
studied fairly extensively
\cite{Sandvik:1995_SqLatt_bilayer,Chubukov:1995_SqLatt_bilayer,Millis:1996_SqLatt_bilayer,Wang:2006_SqLatt_bilayer,Ganesh:2011_honey_bilayer_PRB84}.
As the ratio $J_{1}^{\perp}/J_{1}$ is increased beyond a critical
value $(J_{1}^{\perp}/J_{1})_{c}$ a QPT occurs from a N\'{e}el-ordered
GS phase to a paramagnetic GS phase that is approximately the product
of interlayer dimer valence bonds between NN pairs coupled by
$J_{1}^{\perp}$ bonds.  As we have already noted above, the $T=0$ GS
phase diagram becomes appreciably richer in the additional presence of
frustrating bonds of either the intralayer or interlayer type.  Such
frustrated square-lattice bilayer models have been much studied in
recent years, using a variety of theoretical techniques, both in the
absence \cite{Lin:2000_SqLatt_bilayer,Alet:2016_SqLatt_bilayer} and
presence
\cite{Giamarchi:2008_BoseEinstenCond,Chen:2010_SqLatt_bilayer,Albuquerque:2011_SqLatt_bilayer,Murakami:2013_SqLatt_bilayer,Richter:2006_SqLatt_bilayer,Derzhko:2010_SqLatt_bilayer}
of an external magnetic field.

In the last several years attention has also been paid to analogous
honeycomb-lattice bilayer models, both in the staggered Bernal {\it
  AB} stacking (see, e.g., Ref.\
\cite{Lee:2014_honeycomb_bilayer_ABstack_graphene}) relevant to
bilayer graphene and in the $AA$ stacking (see, e.g., Refs.\
\cite{Ganesh:2011_honey_bilayer_PRB84,Oitmaa:2012_honey_bilayer,Zhang:2014_honey_bilayer,Arlego:2014_honey_bilayer,Bishop:2017_honeycomb_bilayer_J1J2J1perp,Zhang:2016_honey_bilayer,Gomez:2016_honey_bilayer,Krokhmalskii:2017_honey_bilayer})
where the two layers are stacked directly on top of one another.
Since the $AA$ stacking yields the simpler form of coupled-dimer
magnets we restrict attention here to this form of honeycomb bilayer.
After the unfrustrated $J_{1}$--$J_{1}^{\perp}$ honeycomb bilayer was
studied \cite{Ganesh:2011_honey_bilayer_PRB84}, various authors have
studied the effects of both intralayer frustration
\cite{Oitmaa:2012_honey_bilayer,Zhang:2014_honey_bilayer,Arlego:2014_honey_bilayer,Bishop:2017_honeycomb_bilayer_J1J2J1perp}
and interlayer frustration
\cite{Zhang:2016_honey_bilayer,Gomez:2016_honey_bilayer,Krokhmalskii:2017_honey_bilayer}
on the system, by including NNN interactions between spins within the
layers or between the layers, respectively.  In the latter case the
model has been studied both in the absence
\cite{Zhang:2016_honey_bilayer} and presence
\cite{Gomez:2016_honey_bilayer,Krokhmalskii:2017_honey_bilayer} of an
external magnetic field.

There has also been experimental interest in frustrated stacked
honeycomb-lattice bilayer HAFMs.  For example, the Mn$^{4+}$ sites in
the bismuth manganese oxynitrate material
Bi$_{3}$Mn$_{4}$O$_{12}$NO$_{3}$
\cite{Smirnova:2009:honey_spin_3half,Okubo:2010:honey_spin_3half} form
a frustrated spin-$\frac{1}{2}$ $AA$-stacked bilayer honeycomb lattice.  By replacing
the Mn$^{4+}$ ions in this material with V$^{4+}$ ions, it might also
be possible to realize experimentally a spin-$\frac{1}{2}$ HAFM on the
$AA$-stacked honeycomb bilayer.  Ultracold atoms trapped in
optical lattices formed by a periodic potential, which is created by
standing waves formed from a suitable array of lasers, are now also
regularly being used to simulate quantum magnets in a controllable
manner \cite{Bloch:2008_ultraColdAtom_quantumMagnets}.  For example, in the present context, by interfering
three coplanar laser beams propagating at relative angles of
$\pm 120^{\circ}$ one may form a honeycomb lattice \cite{Duan:2003_ultraColdAtom_honeycomb}.  Even more
excitingly, concrete proposals have also been given to form optical
lattices representing honeycomb-lattice bilayers in both $AA$ and
{\it AB} stacking \cite{Tao:2014_ultraColdAtom_honeycomb_bilayer,Dey:2016_ultraColdAtom_honeycomb_bilayer}, using five lasers.

The present study has as one of its goals to investigate the effects
of intralayer frustration on a particularly interesting {\it
  AA}-stacked bilayer honeycomb-lattice version of a
spin-$\frac{1}{2}$ coupled-dimer HAFM that, to our knowledge, has not
been studied before.  In each layer the spins interact via NN, NNN,
and next-next-nearest-neighbor (NNNN) couplings, all of isotropic
Heisenberg type, and with respective exchange constants $J_{1}$,
$J_{2}$, and $J_{3}$.  When all couplings are AFM in nature (i.e.,
$J_{i}>0;\; i=1,2,3$) the classical version of the model (i.e., the
limit $s \rightarrow \infty$) exhibits three phases, viz., two
collinear AFM phases and a spiral (or helical) phase
\cite{Rastelli:1979_honey,Fouet:2001_honey}.  These meet at a
classical triple point located at $J_{3}=J_{2}=\frac{1}{2}J_{1}$.  The
line $J_{3}=J_{2}$ $(\equiv \alpha J_{1})$ is thus of special
interest, and represents the line of maximal frustration in some
sense, which includes the transition point
$\alpha_{{\rm cl}}=\frac{1}{2}$ where the classical GS phase is
macroscopically degenerate.  This $J_{1}$--$J_{2}$--$J_{3}$ model on
the honeycomb lattice has therefore been extensively studied for the
case $s=\frac{1}{2}$, where the effects of quantum fluctuations are
expected to be greatest, especially for the case $J_{3}=J_{2}$ (and
see, e.g., Refs.\
\cite{Cabra:2011_honey,DJJF:2011_honeycomb,Bishop:2012_honey_circle-phase,Li:2012_honey_full}
and references cited therein).  

The plan of the rest of the paper is as follows.  In Sec.\
\ref{model_sec} we describe the model in more detail, including the
known results for the case of vanishing interlayer coupling
($\delta=0$).  To include the interlayer coupling we will use the same theoretical formalism, i.e., the coupled cluster method (CCM), that has been used previously with great success for the corresponding monolayer case.  We will thus briefly review the main elements of the CCM in Sec.\ \ref{ccm_sec} before presenting our
results for the phase boundaries in the $\alpha\delta$ plane of the
two quasiclassical collinear AFM GS phases in Sec.\
\ref{results_sec}.  We conclude with a discussion and summary in Sec.\
\ref{discuss_summary_sec}.

\section{THE MODEL}
\label{model_sec}
\begin{figure*}[t]
\mbox{
\subfigure[]{\includegraphics[width=4.0cm]{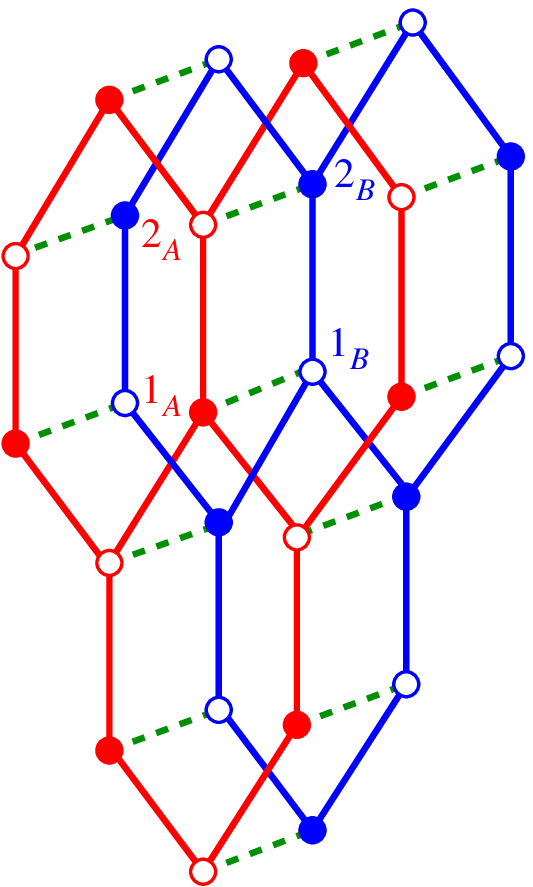}}
\subfigure[]{\includegraphics[width=3.0cm]{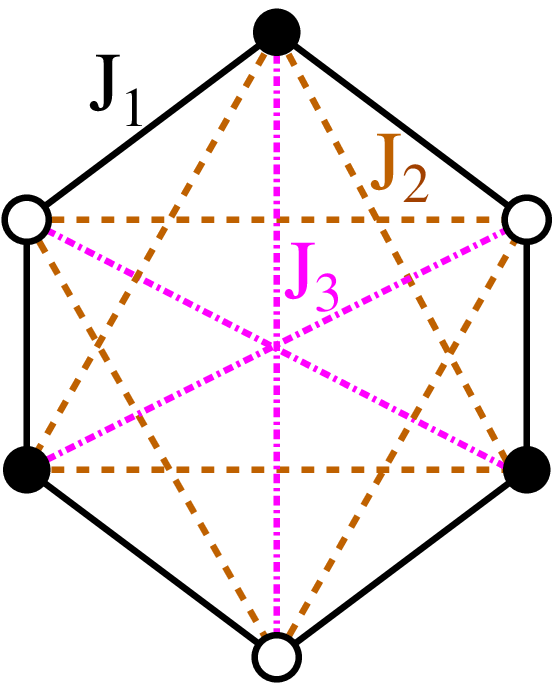}}
\subfigure[]{\includegraphics[width=5.0cm]{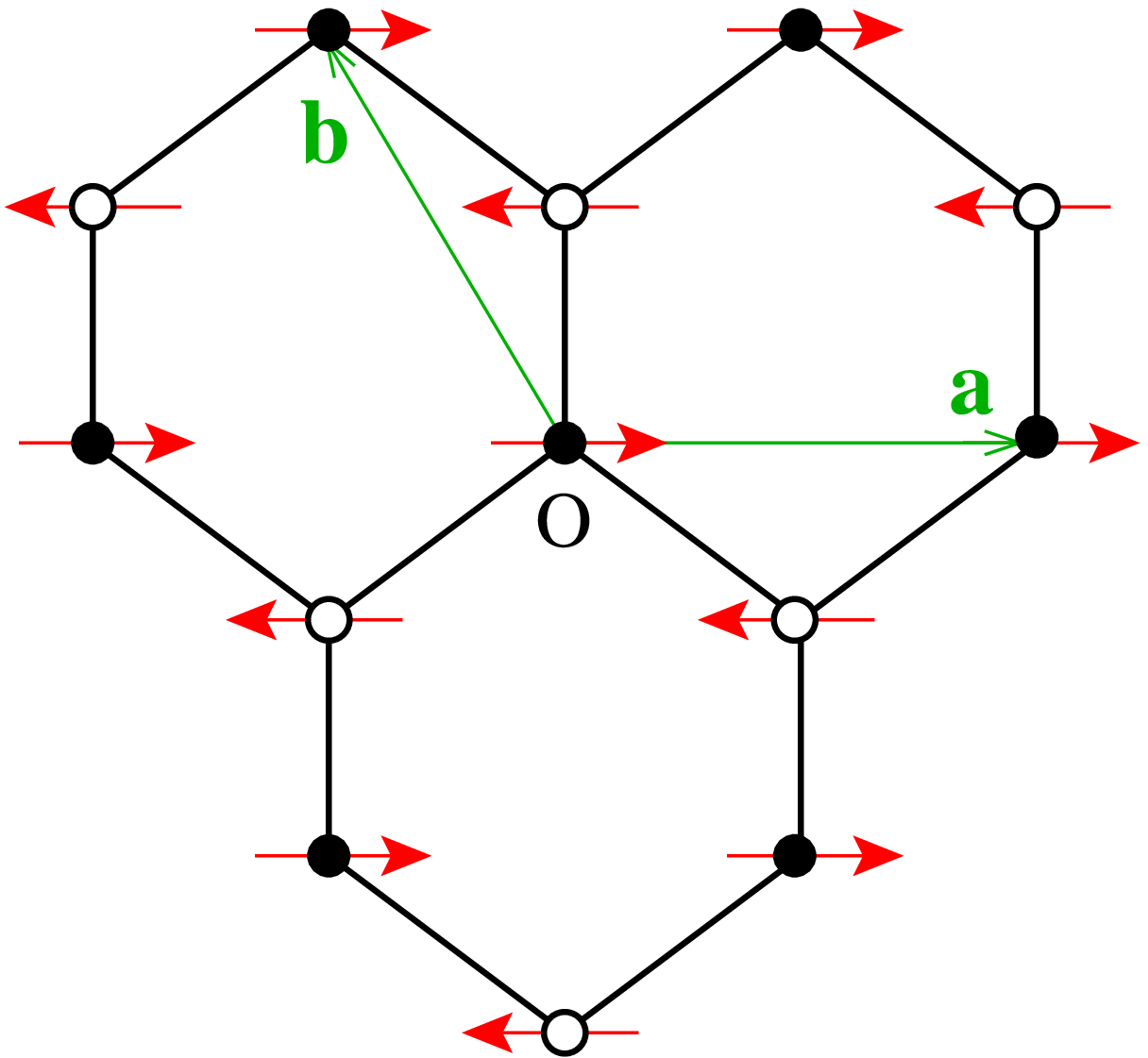}}
\subfigure[]{\includegraphics[width=5.0cm]{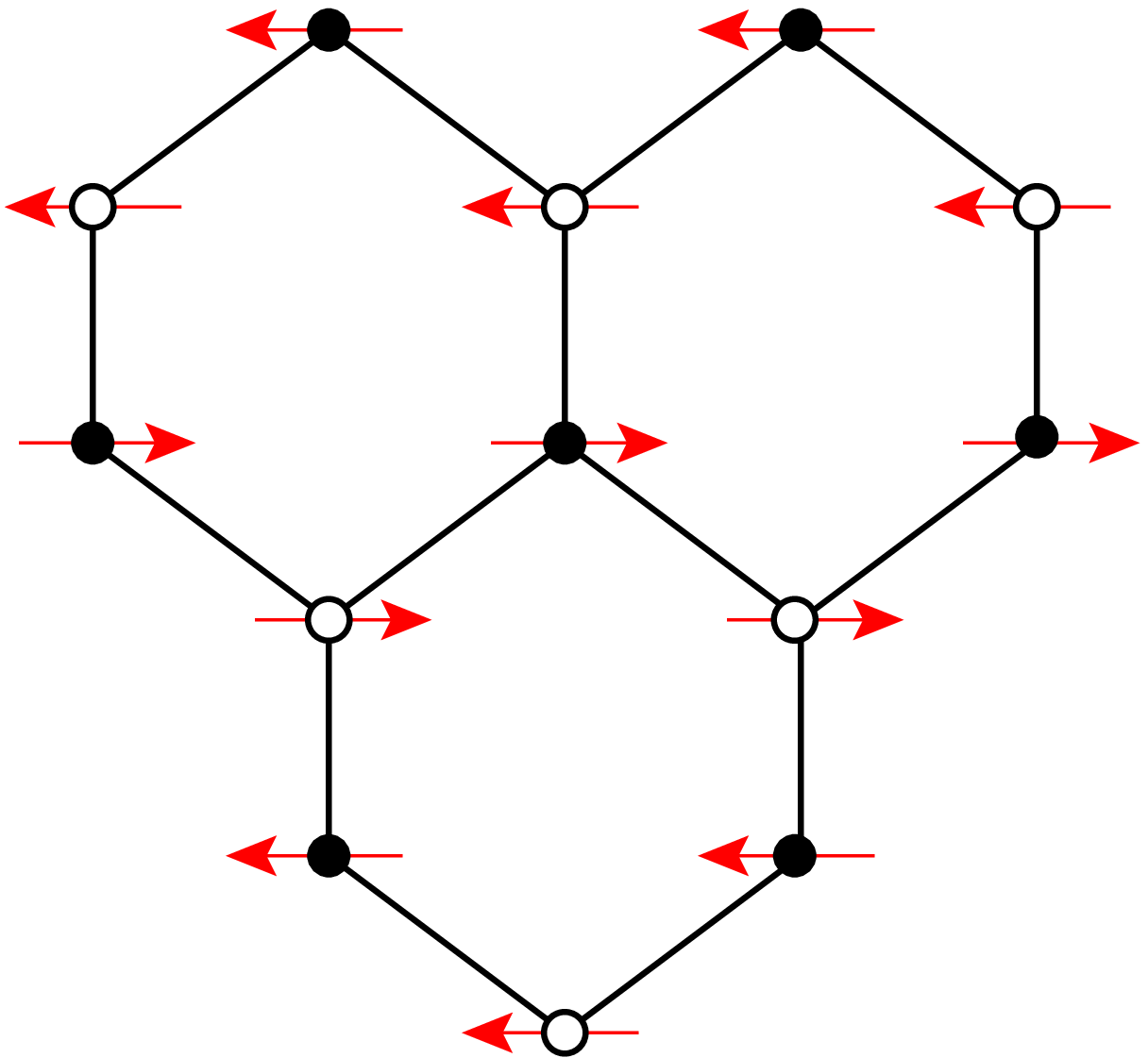}}
}
  \caption{The $J_{1}$--$J_{2}$--$J_{3}$--$J_{1}^{\perp}$ model on the honeycomb bilayer
    lattice, showing (a) the two layers $A$ (red) and $B$ (blue), the nearest-neighbor bonds ($J_{1} = $ -----; $J_{1}^{\perp} = $ - - -)  and the four sites ($1_{A}, 2_{A}, 1_{B}, 2_{B}$) of the unit cell; (b) the intralayer bonds ($J_{1} = $ -----; $J_{2} = $ - - -; $J_{3}=$ $-$$\cdot$$-$$\cdot$$-$) on each monolayer; (c) the triangular Bravais lattice vectors ${\bf a}$ and ${\bf b}$, and the monolayer N\'{e}el state; and (d) one of the three equivalent monolayer striped states.  Sites ($1_{A}$, $2_{B}$) and ($2_{A}$, $1_{B}$) on the two monolayer triangular sublattices are shown by filled and empty circles respectively.}
\label{model_pattern}
\end{figure*}

The $J_{1}$--$J_{2}$--$J_{3}$--$J_{1}^{\perp}$ model on the bilayer
honeycomb lattice is described by the Hamiltonian
\begin{widetext}
\begin{equation}
H=J_{1}{\displaystyle\sum_{{\langle i,j \rangle},\alpha}} \mathbf{s}_{i,\alpha}\cdot\mathbf{s}_{j,\alpha} + 
J_{2}{\displaystyle \sum_{{\langle\langle i,k \rangle\rangle},\alpha}} \mathbf{s}_{i,\alpha}\cdot\mathbf{s}_{k,\alpha} + 
J_{3}{\displaystyle \sum_{{\langle\langle\langle i,l \rangle\rangle\rangle},\alpha}} \mathbf{s}_{i,\alpha}\cdot\mathbf{s}_{l,\alpha}
+ J_{1}^{\perp}{\displaystyle \sum_{i}} \mathbf{s}_{i,A}\cdot\mathbf{s}_{i,B}\,,   \label{H_eq}
\end{equation}
\end{widetext}
where the index $\alpha = A, B$
labels the two layers.  Each site $i$ on each of the two honeycomb
layers carries a spin-$s$ particle denoted by the usual SU(2) spin
operators
${\bf
  s}_{i,\alpha}\equiv(s^{x}_{i,\alpha},s^{y}_{i,\alpha},s^{z}_{i,\alpha})$,
with ${\bf s}^{2}_{i,\alpha} = s(s+1)$, and where for present purposes
we restrict attention to the case $s=\frac{1}{2}$.  In Eq.\ (\ref{H_eq}) the sums over $\langle i,j \rangle$, 
$\langle \langle i,k \rangle \rangle$ and $\langle \langle \langle i,l \rangle \rangle \rangle$ run respectively over
all NN, NNN and NNNN intralayer bonds on each
honeycomb-lattice monolayer, counting each Heisenberg bond once and
once only in each sum.  The last sum in Eq.\ (\ref{H_eq}) describes the interlayer Heisenberg bonds between NN pairs of spins across the two vertically stacked layers.  The pattern of bonds is shown in Figs.\ \ref{model_pattern}(a) and \ref{model_pattern}(b).  
In the present paper we will be interested in the case when all 4
bonds are AFM in nature (i.e., $J_{i}>0,\,i=1,2,3$, and
$J_{1}^{\perp}>0$).  We will also restrict attention, as discussed in
Sec.\ \ref{introd_sec}, to the particularly interesting case when
$J_{3}=J_{2}$.  Since we may regard the exchange coupling constant
$J_{1}$ as simply setting the overall energy scale, the relevant
parameters are thus $(J_{3}/J_{1}=)J_{2}/J_{1}\equiv\alpha$, and
$J_{1}^{\perp}/J_{1}\equiv \delta$.

The honeycomb lattice itself is non-Bravais.  It has a 2-site unit
cell, with two triangular sublattices 1 and 2, and triangular Bravais
lattice vectors ${\mathbf a}$ and ${\mathbf b}$, as shown in Fig.\
\ref{model_pattern}(c).  In terms of unit vectors $\hat{x}$ and
$\hat{z}$ that define the $xz$ plane of the monolayers, we have
${\mathbf a}=\sqrt{3}d\hat{x}$ and
${\mathbf b}=\frac{1}{2}d(-\sqrt{3}\hat{x}+3\hat{z})$, where $d$ is
the NN spacing on the hexagonal lattice.  Sites on sublattice 1 are at
positions ${\mathbf R}_{i}=m{\mathbf a}+n{\mathbf b}$, where
$m,n \in \mathbb{Z}$.  Each unit cell $i$ at position vector
${\mathbf R}_{i}$ on each layer comprises two sites, one at
${\mathbf R}_{i}$ on sublattice 1 and the other at
${\mathbf R}_{i}+d\hat{z}$ on sublattice 2.  In Fig.\
\ref{model_pattern}(a) we also show the corresponding four sites of
the $AA$-stacked bilayer unit cell.

In position space the Wigner-Seitz cell for the monolayer is simply
the parallelogram formed by the triagonal Bravais lattice vectors
${\mathbf a}$ and ${\mathbf b}$.  Equivalently, it may be chosen to be
one of the primitive real-space hexagons of the lattice with side
length $d$, centered on a point of sixfold symmetry.  In that case, in
reciprocal space the first Brillouin zone is then itself also a
hexagon, but which is now rotated by 90$^{\circ}$ with respect to the
real-space Wigner-Seitz hexagon, and with side length
$4\pi/(3\sqrt{3}d)$.  Its first three corners are thus at positions
${\mathbf K}^{(1)}=4\pi/(3\sqrt{3}d)\hat{x}$,
${\mathbf K}^{(2)}=2\pi/(3\sqrt{3}d)\hat{x}+2\pi/(3d)\hat{z}$, and
${\mathbf K}^{(3)}=-2\pi/(3\sqrt{3}d)\hat{x}+2\pi/(3d)\hat{z}$.  The
remaining three corners are at positions
${\mathbf K}^{(i+3)}=-{\mathbf K}^{(i)};\;i=1,2,3$.

Let us first briefly review the situation for the model under
consideration when $\delta=0$ (i.e., for the monolayer).  Along the
line $J_{3}=J_{2}=\alpha J_{1}$ in the classical case there is a
direct transition between the two collinear AFM phases at
$\alpha_{{\rm cl}}=\frac{1}{2}$.  These are the N\'{e}el phase shown
in Fig.\ \ref{model_pattern}(c), which is the GS phase for
$\alpha < \alpha_{{\rm cl}}$, and the striped phase shown in Fig.\
\ref{model_pattern}(d), which is the GS phase for
$\alpha > \alpha_{{\rm cl}}$.  Whereas the N\'{e}el phase on the
honeycomb-lattice monolayer has all three NN pairs of spins
antiparallel to one another at each site, the striped phase has only
one NN pair antiparallel and the other two parallel to one another.
Equivalently, the striped phase is composed of parallel ferromagnetic
zigzag (or sawtooth) chains along one of the three equivalent
honeycomb directions, with neighboring chains coupled
antiferromagnetically.  The striped state shown in Fig.\
\ref{model_pattern}(d) thus has two other equivalent states rotated
with respect to it by $\pm 120^{\circ}$ in the lattice plane.

As usual, the classical phases may generically be described by an
ordering wave vector ${\mathbf Q}$, together with a parameter
$\theta$ that measures the angle between the two spins in each
monolayer unit cell $i$ at position vector ${\mathbf R}_{i}$.  The
classical spins, of length $s$, are thus written as
\begin{equation}
{\mathbf s}_{i,\rho}=s[\cos({\mathbf Q}\cdot{\mathbf R}_{i}+\theta_{\rho})\hat{x}_{s}+\sin({\mathbf Q}\cdot{\mathbf R}_{i}+\theta_{\rho})\hat{z}_{s}]\,,
\end{equation}
where the index $\rho$ labels the two sites in the unit cell, and
$\hat{x}_{s}$ and $\hat{z}_{s}$ are two orthogonal unit vectors that define
the plane of the spins.  The angles $\theta_{\rho}$ may be chosen with
no loss of generality so that $\theta_{1}=0$ for spins on
triangular-sublattice 1 and $\theta_{2}=\theta$ for spins on
triangular-sublattice 2.  In this description of the classical phases
the N\'{e}el phase shown in Fig.\ \ref{model_pattern}(c) has wave
vector ${\mathbf Q}=0$ and $\theta=\pi$.  Similarly, the striped phase
shown in Fig.\ \ref{model_pattern}(d) has wave vector
${\mathbf Q}={\mathbf M}^{(2)}$ and $\theta=\pi$, where
${\mathbf M}^{(2)}=2\pi/(3d)\hat{z}$ is the vector of the midpoint of
the edge joining the corners of the hexagonal first Brillouin zone at
positions ${\mathbf K}^{(2)}$ and ${\mathbf K}^{(3)}$.  The two other
inequivalent striped states are hence described by the wave vectors of
the remaining inequivalent midpoints of the other two edges of the
first Brillouin zone, and in each case now with $\theta=0$.  Thus, the
other two striped states have wave vectors
${\mathbf Q} = {\mathbf M}^{(1)}
=\pi/(\sqrt{3}d)\hat{x}+\pi/(3d)\hat{z}$ (that is the midpoint of the
edge joining corners ${\mathbf K}^{(1)}$ and ${\mathbf K}^{(2)}$), and
${\mathbf Q} = {\mathbf M}^{(3)}
=-\pi/(\sqrt{3}d)\hat{x}+\pi/(3d)\hat{z}$ (that is the midpoint of the
edge joining corners ${\mathbf K}^{(3)}$ and ${\mathbf K}^{(4)}$).

Let us now compare this classical result for the monolayer
($\delta=0$) version of our model with the corresponding case when
$s=\frac{1}{2}$.  In the spin-$\frac{1}{2}$ case the classical
transition point is split into two quantum critical points (QCPs) at
$\alpha_{c_{1}} < \alpha_{{\rm cl}}$ and
$\alpha_{c_{2}} > \alpha_{{\rm cl}}$, with a magnetically disordered
paramagnetic GS phase in the intermediate region
\cite{Cabra:2011_honey,DJJF:2011_honeycomb}.  Lowest-order spin-wave
theory, for example, provides the estimates \cite{Cabra:2011_honey}
$\alpha_{c_{1}} \approx 0.29$ and $\alpha_{c_{2}} \approx 0.55$.  By
contrast, the more powerful, and potentially more accurate, method of
Schwinger-boson mean-field theory (SBMFT) yields the estimates
\cite{Cabra:2011_honey} $\alpha_{c_{1}} \approx 0.41$ and
$\alpha_{c_{2}} \approx 0.6$.  SBMFT also predicts a quantum
disordered phase in the intermediate regime
$\alpha_{c_{1}} < \alpha < \alpha_{c_{2}}$, where a gap opens up in
the bosonic dispersion and the spin-spin correlation function displays
traces of N\'{e}el short-range order.  These results are broadly
confirmed by high-order CCM calculations \cite{DJJF:2011_honeycomb},
which yield the values $\alpha_{c_{1}} \approx 0.47$ and
$\alpha_{c_{2}} \approx 0.60$.  Furthermore, CCM calculations of the
plaquette valence-bond crystalline (PVBC) susceptibility
\cite{DJJF:2011_honeycomb} provide strong evidence for the intermediate
paramagnetic phase to be a gapped state with PVBC order over the
entire region.

Of special interest for the spin-$\frac{1}{2}$ monolayer, the QPT at
$\alpha_{c_{1}}$ between the N\'{e}el phase (that is the stable GS
phase for $\alpha < \alpha_{c_{1}}$) and the paramagnetic phase (that
is the stable GS phase for $\alpha_{c_{1}} < \alpha < \alpha_{c_{2}}$)
appears to be continuous, while that at $\alpha_{c_{2}}$ appears to be
of first-order type \cite{DJJF:2011_honeycomb}.  Since the N\'{e}el
and intermediate phases break different symmetries, the QPT at
$\alpha_{c_{1}}$ is thus favored to be described by the scenario of
deconfined quantum criticality
\cite{Senthil:2004_Science_deconfinedQC_merge,*Senthil:2004_PRB_deconfinedQC_merge}.
In view of this rich scenario it seems of considerable interest to
study the comparable spin-$\frac{1}{2}$
$J_{1}$--$J_{2}$--$J_{3}$--$J_{1}^{\perp}$ model on the $AA$-stacked
honeycomb bilayer, where we now include AFM interlayer NN bonds of
strength $J_{1}^{\perp} > 0$.  Once again we will study the model here
along the line $J_{3}=J_{2}\equiv \alpha J_{1}$ with
$J_{1}^{\perp} \equiv \delta J_{1}$.  Specific goals will be to study
how the QCPs $\alpha_{c_{1}}$ and $\alpha_{c_{2}}$ now depend on the
interlayer coupling parameter $\delta$.  To that end we will use the
same theoretical technique, viz., the coupled cluster method (CCM), as
has been used previously to describe accurately the corresponding
monolayer case ($\delta=0$) \cite{DJJF:2011_honeycomb}.

Let us now turn to the corresponding case of the
$J_{1}$--$J_{2}$--$J_{3}$--$J_{1}^{\perp}$ model (with $J_{3}=J_{2}$),
on the $AA$-stacked honeycomb-lattice bilayer, which we aim to study
here for the spin-$\frac{1}{2}$ case.  At the classical
$(s \rightarrow \infty)$ level the inclusion of an NN interlayer
coupling with strength $J_{1}^{\perp}>0$ introduces no extra
frustration, and its effect is essentially trivial.  Simply the NN
interlayer pairs of spins anti-align, with each monolayer having the same N\'{e}el order for
$\alpha < \alpha_{{\rm cl}}$ or striped order for
$\alpha > \alpha_{{\rm cl}}$ as in the absence of interlayer pairing.
By contrast, as we have discussed in detail in Sec.\ \ref{introd_sec},
the spin-$\frac{1}{2}$ case is expected to be of greater interest and
subtlety, due to the expected formation of NN interlayer spin-singlet
dimers in the large-$\delta$ limit, where the GS phase will thus be a
valence-bond crystalline (VBC) phase formed from interlayer dimers.
This interlayer dimer VBC (IDVBC) phase will be gapped, by contrast to the
gapless nature of the quasiclassical N\'{e}el and striped phases with
magnetic LRO, where the magnon excitations are massless Goldstone
modes.  In the complete IDVBC phase (i.e., in the limit
$\delta \rightarrow \infty$) the lowest-lying excited state is a spin-1
state formed by breaking a single NN interlayer dimer from a
spin-singlet to a spin-triplet state.  Thus, we expect the scaled
triplet spin gap to be given, in the limiting case, by
\begin{equation}
\frac{\Delta}{J_{1}} \xrightarrow[\delta \to \infty]{} \delta\,. \label{triplet_spin_gap_scaled}
\end{equation}

As we have noted above, the phase diagram of the model along the line
$\delta=0$ (i.e., for the monolayer) is already a rich one, with two
QCPs in the range $0 \leq \alpha \leq 1$ of the frustration parameter,
with one being continuous (and hence probably of a deconfined quantum
critical nature) and the other being of first-order type.  Clearly,
the enlargement of the model by the addition of the extra parameter
$\delta$ can only increase our understanding of the quantum phase
diagram, even for the case of the monolayer.  In particular, the two
QCPs for the limiting case $\delta=0$ now become quantum critical
lines (or quantum phase boundary lines) in the $\alpha\delta$ plane.
As a foretaste of our overall results, one of our most important
findings is that these two phase boundary lines display a marked
``avoided crossing'' behavior, with both curves consequently
exhibiting a distinct reentrant nature.  This finding by itself
clearly throws new light on the phase diagram of the monolayer
($\delta=0$), as discussed more fully in Secs.\ \ref{results_sec} and
\ref{discuss_summary_sec}.

In order to examine the effect of interlayer coupling on the QCPs
$\alpha_{c_{1}}$ and $\alpha_{c_{2}}$ of the spin-$\frac{1}{2}$
honeycomb-lattice monolayer we will therefore present in Sec.\
\ref{results_sec} results for both the magnetic order parameter (i.e.,
the average local on-site GS magnetization) $M$ and the triplet spin
gap $\Delta$, for each of the N\'{e}el and striped quasiclassical
phases, as functions of both parameters $\alpha$ and $\delta$.  We
utilize both perfectly-ordered states as model wave functions, on top
of which we include quantum fluctuations in a fully systematic
formalism (viz., the CCM), as we first demonstrate in Sec.\
\ref{ccm_sec}.  In particular we show how the CCM can be implemented
in very high orders in a well-defined and fully systematic
approximation hierarchy, to yield sequences of approximants for both
$M$ and $\Delta$.  We show further how these sequences can then also
be extrapolated, in a controlled and stable manner, to the limit where
the corresponding wave functions are exact in principle.  These
extrapolations are the {\it only} approximations made.  In Sec.\
\ref{results_sec} we will show explicitly how the results for both $M$
and $\Delta$ yield accurate and consistent estimates of the phase
boundaries, $\alpha_{c_{1}}(\delta)$ and $\alpha_{c_{2}}(\delta)$, of
the N\'{e}el and striped GS phases.

\section{THE COUPLED CLUSTER METHOD}
\label{ccm_sec}
The CCM
\cite{Kummel:1978_ccm,Bishop:1978_ccm,Bishop:1982_ccm,Arponen:1983_ccm,Bishop:1987_ccm,Arponen:1987_ccm,Bartlett:1989_ccm,Arponen:1991_ccm,Bishop:1991_TheorChimActa_QMBT,Bishop:1998_QMBT_coll,Zeng:1998_SqLatt_TrianLatt,Fa:2004_QM-coll}
is one of the very few size-extensive and size-consistent techniques
of quantum many-body theory.  It thereby provides results in the
$N \rightarrow \infty$ limit (where $N$ is the number of particles,
i.e., lattice spins in our case) from the outset, at {\it all} levels
of approximation.  Hence, no finite-size scaling is ever required.
Particularly apposite to the CCM also is the fact that both the
Goldstone linked-cluster theorem and the very important
Hellmann-Feynman theorem are also preserved at {\it every} level of
approximate implementation of the formalism.  The latter plays a large
part in ensuring that the method yields accurate, self-consistent, and
robust results for a variety of physical parameters for a given
system.  The method has been applied very widely, yielding results of
great (and often unsurpassed) accuracy to systems as diverse as finite
nuclei \cite{Kummel:1978_ccm}, the electron gas (or jellium)
\cite{Bishop:1978_ccm,Bishop:1982_ccm}, atoms and molecules of
interest in quantum chemistry \cite{Bartlett:1989_ccm}, and a broad
spectrum of spin-lattice problems of interest in quantum magnetism
\cite{DJJFarnell:2014_archimedeanLatt,Bishop:2017_honeycomb_bilayer_J1J2J1perp,Zeng:1998_SqLatt_TrianLatt,Fa:2004_QM-coll,Bishop:1994_ccm_XXZ_SqLatt,Zeng:1995_ccm_triangLatt,Zeng:1996_SqLatt_TrianLatt,Bishop:2000_XXZ,Kruger:2000_JJprime,Fa:2001_SqLatt_s1,Darradi:2005_Shastry-Sutherland,Bi:2008_EPL_J1J1primeJ2_s1,Bi:2008_JPCM_J1xxzJ2xxz_s1,Bi:2009_SqTriangle,Bishop:2010_UJack,Bishop:2010_KagomeSq,Bishop:2011_UJack_GrtSpins,PHYLi:2012_SqTriangle_grtSpins,PHYLi:2012_honeycomb_J1neg,Li:2012_anisotropic_kagomeSq,Bishop:2012_honeyJ1-J2,RFB:2013_hcomb_SDVBC,Richter:2015_ccm_J1J2sq_spinGap,Bishop:2015_J1J2-triang_spinGap,Bishop:2015_honey_low-E-param,Bishop:2016_honey_grtSpins,Li:2016_honey_grtSpins,Li:2016_honeyJ1-J2_s1,Li:2018_crossStripe_low-E-param}.

To initiate the CCM in practice one needs to choose a suitable model
(or reference) state to act as a generalized vacuum state.  The
quantum correlations present in the exact GS or excited-state (ES)
wave function of the system are then systematically incorporated on
top of the model state in a hierarchical scheme that becomes exact in
some limit, which is usually unattainable at particular levels of
computational implementation.  Appropriate conditions for a state to
be a suitable CCM model state have been discussed extensively in the
literature
\cite{Arponen:1983_ccm,Arponen:1987_ccm,Bishop:1991_TheorChimActa_QMBT,Bishop:1998_QMBT_coll,Zeng:1998_SqLatt_TrianLatt,Fa:2004_QM-coll}.
For spin-lattice models, however, {\it all} (quasi)classical states
with perfect magnetic LRO are suitable CCM model states.  Hence, we
use here both the N\'{e}el and striped states shown in Figs.\
\ref{model_pattern}(c) and \ref{model_pattern}(d) respectively, for
each honeycomb-lattice monolayer, in each case with NN interlayer
spins on the $AA$-stacked bilayer aligned antiparallel to each other,
as our two choices for CCM model state.  We will present independent
sets of results in Sec.\ \ref{results_sec} based on both model states
taken separately.

We shall only briefly review here some of the principal features and
most important elements of the CCM, and refer the reader to Refs.\
\cite{Bishop:2017_honeycomb_bilayer_J1J2J1perp,Kummel:1978_ccm,Bishop:1978_ccm,Bishop:1982_ccm,Arponen:1983_ccm,Bishop:1987_ccm,Arponen:1987_ccm,Bartlett:1989_ccm,Arponen:1991_ccm,Bishop:1991_TheorChimActa_QMBT,Bishop:1998_QMBT_coll,Zeng:1998_SqLatt_TrianLatt,Fa:2004_QM-coll}
for a fuller description.  It is very convenient to be able to treat
each lattice spin in each model state as being equivalent to one
another.  A simple way to do so is to perform a separate passive
rotation of each such spin so that they all point in the same direction,
say downwards (i.e., along the local negative $z_{s}$ axis), in its
own local spin-coordinate frame.  Accordingly, after such a choice of
local spin-coordinate frames has been made, each model state takes the
universal form
$|\Phi\rangle =
|\downarrow\downarrow\downarrow\cdots\downarrow\rangle$.  Naturally,
the Hamiltonian also needs to be rewritten appropriately for each such
choice.

For a completely general quantum many-body system, its exact GS energy
eigenket $|\Psi\rangle$, where $H|\Psi\rangle = E|\Psi\rangle$, is
expressed in the exponentiated form,
\begin{equation}
|\Psi\rangle = {\rm e}^{S}|\Phi\rangle\,; \quad S=\sum_{I\neq 0}{\cal{S}}_{I}C_{I}^{+}\,,  \label{ket_parametrization_eq}
\end{equation}
that is a hallmark of the CCM.  Its GS energy eigenbra counterpart
$\langle\tilde{\Psi}|$, where
$\langle\tilde{\Psi}|H = E\langle\tilde{\Psi}|$, is correspondingly
expressed in the CCM parametrization,
\begin{equation}
\langle\tilde{\Psi}|=\langle\Phi|\tilde{S}{\rm e}^{-S}\,;  \quad \tilde{S}=1 + \sum_{I \neq 0} {\cal{\tilde{S}}}_{I}C_{I}^{-}\,,  \label{bra_parametrization_eq}
\end{equation}
where $C_{I}^{-} \equiv (C_{I}^{+})^{\dagger}$, and
$C_{0}^{+} \equiv 1$, the identity operator.  The set-index $I$ here
represents a multiparticle configuration, such that the set of states
$\{C_{I}^{+}|\Phi\rangle\}$ completely spans the many-body Hilbert space.  The
model state $|\Phi\rangle$ and the complete set of mutually commuting,
multiconfigurational creation operators $\{C_{I}^{+}\}$ must be chosen
so that the former is a fiducial vector (or generalized vacuum state)
with regard to the latter, in the sense that they obey the conditions,
\begin{equation}
\langle\Phi|C_{I}^{+} = 0 = C_{I}^{-}|\Phi\rangle\,, \quad \forall I \neq 0\,,
\end{equation}
as well as
\begin{equation}
[C_{I}^{+},C_{J}^{+}]=0\,, \quad \forall I,J\,.  \label{creation_operator_commutation_relation}
\end{equation}

The model state $|\Phi\rangle$ is chosen to be normalized,
$\langle\Phi|\Phi\rangle\equiv 1$, and the CCM parametrization of Eq.\
(\ref{ket_parametrization_eq}) ensures that the exact GS wave function
$|\Psi\rangle$ obeys the intermediate normalization condition,
$\langle\Phi|\Psi\rangle=1$.  Similarly, the CCM parametrization of
Eq.\ (\ref{bra_parametrization_eq}) for $\langle\tilde{\Psi}|$ ensures
  the automatic fulfillment of the normalization condition
  $\langle\tilde{\Psi}|\Psi\rangle = 1$.  In practice, it is also
  convenient to orthonormalize the set of states
  $\{C_{I}^{+}|\Phi\rangle\}$, i.e., so that they obey the relations,
\begin{equation}
\langle\Phi|C_{I}^{-}C_{J}^{+}|\Phi\rangle=\delta_{I,J}\,, \quad \forall I,J \neq 0\,,  \label{create_destruct_operators_orthonornmal_Eq}
\end{equation}
where $\delta_{I,J}$ is a suitably generalized Kronecker symbol.

We note that Hermiticity clearly ensures that the destruction correlation operator $\tilde{S}$ may formally be expressed in terms of its creation counterpart $S$ via the relation
\begin{equation}
\langle\Phi|\tilde{S} = \frac{\langle\Phi|{\rm e}^{S^{\dagger}}{\rm e}^{S}}{\langle\Phi|{\rm e}^{S^{\dagger}}{\rm e}^{S}|\Phi\rangle}\,.  \label{Destruct_operator_hermiticity_Eq}
\end{equation}
A key feature of the CCM is that the constraint of Eq.\
(\ref{Destruct_operator_hermiticity_Eq}) is not imposed explicitly.
Rather, the operator $\tilde{S}$ is formally decomposed independently of $S$,
as in Eq.\ (\ref{bra_parametrization_eq}).  Naturally, in the exact
limit, when all multiconfigurational clusters specified by the
complete set $\{I\}$ are retained, Eq.\
(\ref{Destruct_operator_hermiticity_Eq}) would be exactly fulfilled.
In practice, when approximations are made to restrict the set $\{I\}$
to some manageable subset, Eq.\
(\ref{Destruct_operator_hermiticity_Eq}) may only be approximately
fulfilled.  This manifest loss of Hermiticity, however, is more than
compensated by the gain that the Hellmann-Feynman theorem is itself
exactly fulfilled by the CCM parametrizations of Eqs.\
(\ref{ket_parametrization_eq}) and (\ref{bra_parametrization_eq}),
even when the sums over the multiconfigurational indices $\{I\}$ are
restricted.

Turning now to the specific case of a quantum spin-lattice system, in
the local spin-coordinate frames discussed above, where the CCM model
state takes the universal form
$|\Phi\rangle =
|\downarrow\downarrow\downarrow\cdots\downarrow\rangle$, the operator
$C_{I}^{+}$ can now also be chosen to have the universal form of a
product of single-spin raising operators,
$s_{k}^{+}\equiv s_{k}^{x}+is_{k}^{y}$.  Thus, the set-index $I$ now takes
the form of a set of site indices,
\begin{equation}
I \equiv \{k_{1},k_{2},\cdots,k_{n}\,; \quad n=1,2\cdots 2sN\}\,,
\end{equation}
in which no given site index $k_{i}$ may appear more than 2$s$ times.  Correspondingly, the operator $C_{I}^{+}$ creates a multispin configuration cluster,
\begin{equation}
C^{+}_{I} \equiv s^{+}_{k_{1}}s^{+}_{k_{2}}\cdots s^{+}_{k_{n}};\; \quad n=1,2,\cdots 2sN\,.
\end{equation}

Clearly, all the GS quantities may now be expressed wholly in terms of
the CCM correlation coefficients
$\{{\cal S}_{I},\tilde{{\cal S}}_{I}\}$.  For example, the GS magnetic
order parameter, which is simply the average local on-site
magnetization, takes the form
\begin{equation}
M = -\frac{1}{N}\langle\Phi|\tilde{S}\sum_{k=1}^{N}{\rm e}^{-S}s_{k}^{z}{\rm e}^{S}|\Phi\rangle\,,  \label{M_definition_eq}
\end{equation}
where $s_{k}^{z}$ is expressed in the local rotated spin axes
described above.  Thus, all that remains for the GS calculations is to
calculate the coefficients $\{{\cal S}_{I}, \tilde{{\cal S}}_{I}\}$.

Formally, this is done
by minimizing the GS energy expectation functional,
\begin{equation}
\bar{H}=\bar{H}[{\cal S}_{I},{\tilde{\cal S}_{I}}]\equiv
\langle\Phi|\tilde{S}{\rm e}^{-S}H{\rm e}^{S}|\Phi\rangle\,,  
\end{equation}
with respect to all coefficients
$\{{\cal S}_{I},{\tilde{\cal S}}_{I}\,; \forall I \neq 0\}$.
Extremization with respect to $\tilde{{\cal S}}_{I}$, using Eq.\
(\ref{bra_parametrization_eq}), trivially yields the relations
\begin{equation}
\langle\Phi|C^{-}_{I}{\rm e}^{-S}H{\rm e}^{S}|\Phi\rangle = 0\,, \quad \forall I \neq 0\,,  \label{non_linear_ket_Eq}
\end{equation}
which are a coupled set of nonlinear equations for the coefficients
$\{{\cal S}_{I}\}$.  Similarly, use of Eq.\
(\ref{ket_parametrization_eq}) and extremization of $\bar{H}$ with
respect to ${\cal S}_{I}$, yields the relations
\begin{equation}
\langle\Phi|\tilde{S}({\rm e}^{-S}H{\rm e}^{S}-E)C^{+}_{I}|\Phi\rangle=0\,, \quad \forall I \neq 0\,,   \label{linear_bra_Eq_equivalentForm}
\end{equation}
where we have also used the simple relation $[S,C_{I}^{+}]=0$, which
follows from the explicit CCM parametrization of Eq.\
(\ref{ket_parametrization_eq}) together with Eq.\
(\ref{creation_operator_commutation_relation}).  Equation
(\ref{linear_bra_Eq_equivalentForm}) is just a set of linear equations
for the coefficients $\{\tilde{{\cal S}}_{I}\}$, once the coefficients
$\{{\cal S}_{I}\}$ are input, having first been obtained by solving
Eq.\ (\ref{non_linear_ket_Eq}).

An ES energy eigenket $|\Psi_{e}\rangle$, where $H|\Psi_{e}\rangle=E_{e}|\Psi_{e}\rangle$ is similarly expressed in the CCM formalism in terms of a linear excitation operator, $X^{e}$, as
\begin{equation}
|\Psi_{e}\rangle = X^{e}{\rm e}^{S}|\Phi\rangle\,; \quad X^{e} = \sum_{I \neq 0}{\cal X}_{I}^{e}C_{I}^{+}\,,  \label{excited_ket_parametrization_eq}
\end{equation}
as the analog of Eq.\ (\ref{ket_parametrization_eq}) for the GS
counterpart.  Using the obvious commutativity relation, $[X^{e},S]=0$,
which follows from Eqs.\ (\ref{ket_parametrization_eq}),
(\ref{creation_operator_commutation_relation}), and
(\ref{excited_ket_parametrization_eq}), it is easy to combine the GS
and ES Schr\"{o}dinger equations to yield the equation
\begin{equation}
e^{-S}[H,X^{e}]e^{S}|\Phi\rangle = \Delta_{e}X^{e}|\Phi\rangle\,,   \label{excited_state_Eq}
\end{equation}
where $\Delta_{e}$ is the excitation energy,
\begin{equation}
\Delta_{e} \equiv E_{e}-E\,.   \label{excited_energy_Eq}
\end{equation}
By taking the inner product of Eq.\ (\ref{excited_state_Eq}) with
$\langle\Phi|C_{I}^{-}$, and making use of Eq.\
(\ref{create_destruct_operators_orthonornmal_Eq}), one may readily
derive the set of equations,
\begin{equation}
\langle\Phi|C_{I}^{-}({\rm e}^{-S}H{\rm e}^{S}-E)X^{e}|\Phi\rangle = \Delta_{e}{\cal X}_{I}^{e}\,, \quad \forall I \neq 0\,.  \label{excited_state_coeff_Eq}
\end{equation}
Once the operator $({\rm e}^{-S}H{\rm e}^{S}-E)$ has been input into
Eq.\ (\ref{excited_state_coeff_Eq}) from the solution to Eq.\
(\ref{non_linear_ket_Eq}), Eq.\ (\ref{excited_state_coeff_Eq}) is
simply a set of generalized linear eigenvalue equations for the ES
ket-state CCM correlation coefficients $\{{\cal X}_{I}^{e}\}$ and the
excitation energy (eigenvalue) $\Delta_{e}$.

Everything so far has been formally exact, and we turn now to where
approximations may be needed for computational implementation of the
CCM formalism.  One possible source of approximation could involve the
evaluation of the exponentiated operators ${\rm e}^{\pm S}$ that lie
at the heart of the CCM parametrizations of Eqs.\
(\ref{ket_parametrization_eq}), (\ref{bra_parametrization_eq}), and
(\ref{excited_ket_parametrization_eq}).  However, we note that in each
of Eqs.\ (\ref{non_linear_ket_Eq}),
(\ref{linear_bra_Eq_equivalentForm}), and
(\ref{excited_state_coeff_Eq}) that need to be solved for the CCM
correlation coefficients $\{{\cal S}_{I}\}$,
$\{{\tilde{\cal S}}_{I}\}$ and $\{{\cal X}_{I}^{e}\}$, these appear
only in the combination ${\rm e}^{-S}H{\rm e}^{S}$ of a similarity transform of
the system Hamiltonian.  We have described in detail elsewhere (and
see, e.g., Refs.\
\cite{Fa:2004_QM-coll,Bishop:2015_honey_low-E-param,Li:2016_honeyJ1-J2_s1}
and references cited therein) how the otherwise infinite-order nested
commutator expansion
\begin{equation}
{\rm e}^{-S}H{\rm e}^{S}=\sum_{n=0}^{\infty}\frac{1}{n!}[H,S]_{n}\,,  \label{H_similarity_transform_expansion_Eq}
\end{equation}
where $[H,S]_{n}$, the $n$-fold nested commutator, is defined
iteratively  as
\begin{equation}
[H,S]_{n} \equiv [[H,S]_{n-1},S]\,; \quad [H,S]_{0}=H\,,
\end{equation}
actually terminates exactly in the present case at the term with
$n=2$.  Similarly, all GS expectation values, such as the magnetic
order parameter $M$ of Eq.\ (\ref{M_definition_eq}) may be evaluated
without any truncations of the similarity transform
${\rm e}^{-S}s_{k}^{z}{\rm e}^{S}$.

Hence, the sole approximation made is in the choices of which subsets
of multispin-flip configurations $\{I\}$ to retain in the expansions
of correlation coefficients for both the GS wave functions, as in Eqs.\
(\ref{ket_parametrization_eq}), (\ref{bra_parametrization_eq}), and
the ES wave function considered, as in Eq.\
(\ref{excited_ket_parametrization_eq}).  In the present case we
restrict attention to the lowest-lying spin-triplet excitation, so
that $\Delta_{e}$ now becomes equal to the spin-triplet gap, which we
denote by $\Delta$.  For both the GS and ES calculations we employ the
well-known lattice animal-based subsystem (LSUB$n$) truncation
hierarchy, which has been much studied and utilized for a wide variety
of spin-lattice problems in the past (and see e.g., Refs.\
\cite{DJJFarnell:2014_archimedeanLatt,Bishop:2017_honeycomb_bilayer_J1J2J1perp,DJJF:2011_honeycomb,Li:2012_honey_full,Zeng:1998_SqLatt_TrianLatt,Fa:2004_QM-coll,Li:2016_honeyJ1-J2_s1}
and references cited therein).

At the LSUB$n$ level of approximation one retains all multispin-flip
configurations $I$ in the CCM correlation operators ${\cal S}$,
$\tilde{{\cal S}}$ and ${\cal X}^{e}$ defined in Eqs.\
(\ref{ket_parametrization_eq}), (\ref{bra_parametrization_eq}), and
(\ref{excited_ket_parametrization_eq}), respectively, which are
restrained to all distinct locales on the lattice that contain no more
than $n$ contiguous sites.  A cluster of sites is said to be
contiguous (or to form a lattice animal or polyomino) when each site
of the cluster is NN to at least one other, taking into account the
geometrical definition of NN sites employed.  For the GS calculation
we restrict the multispin-flip configurations at each LSUB$n$ level to
have $s_{T}^{z}=0$, where $s_{T}^{z} \equiv \sum_{k=1}^{N}s_{k}^{z}$,
defined in the global spin axes (i.e., {\it before} the local rotations
discussed above have been performed).  Similarly, for the ES
calculation of the triplet spin gap $\Delta$, we restrict the
comparable configurations to have $s_{T}^{z}=1$.  In both cases we
utilize the space- and point-group symmetries of the Hamiltonian and
the CCM model state $|\Phi\rangle$ being employed to reduce the number
of fundamentally distinct configurations to a minimum, $N_{f}(n)$.
Since the number $N_{f}(n)$ increases rapidly as a function of
truncation index $n$, it soon becomes necessary to use supercomputing
resources and massive parallelization, together with custom-made
computer-algebraic packages, to enumerate the independent cluster
configurations retained and to derive the corresponding CCM equations
at a given LSUB$n$ level, and then finally to solve them
\cite{Zeng:1998_SqLatt_TrianLatt,ccm_code}.  For the present bilayer
model we are thereby able to perform both the GS and ES calculations
reported in Sec.\ \ref{results_sec} at LSUB$n$ levels of approximation
with $n \leq 10$.  For the GS calculation we have $N_{f}(10)=70\,118$
$(175\,223)$ using the bilayer N\'{e}el (striped) states described
previously as CCM model states.  For the corresponding ES calculation
of $\Delta$ we have $N_{f}(10) = 121\,103$ $(320\,476)$ using the
bilayer N\'{e}el (striped) states as CCM model states.  Both numbers
are larger for the striped state than the N\'{e}el state since the
former has less symmetries than the latter.  We note that for this
model the geometric definition of contiguous sites for the LSUB$n$
configurations simply corresponds to NN pairs connected by $J_{1}$ or
$J_{1}^{\perp}$ bonds.

\begin{figure*}[t]
\mbox{
\hspace{-1.0cm}
\subfigure[]{\includegraphics[width=7.5cm]{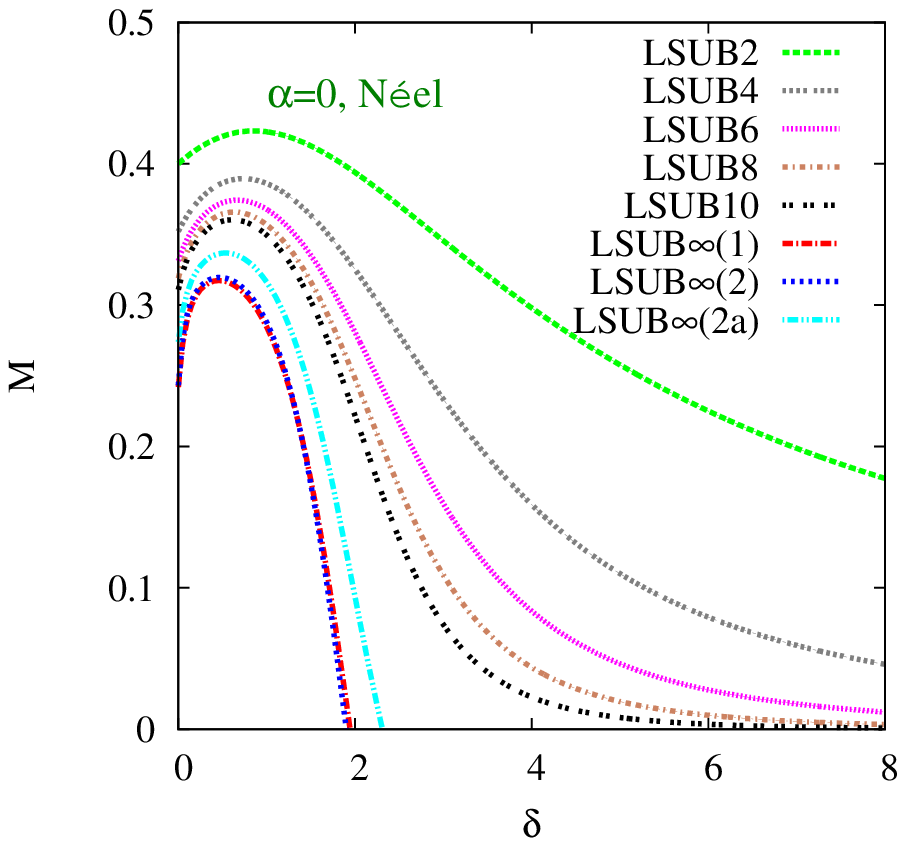}}
\hspace{-1.8cm}
\subfigure[]{\includegraphics[width=7.5cm]{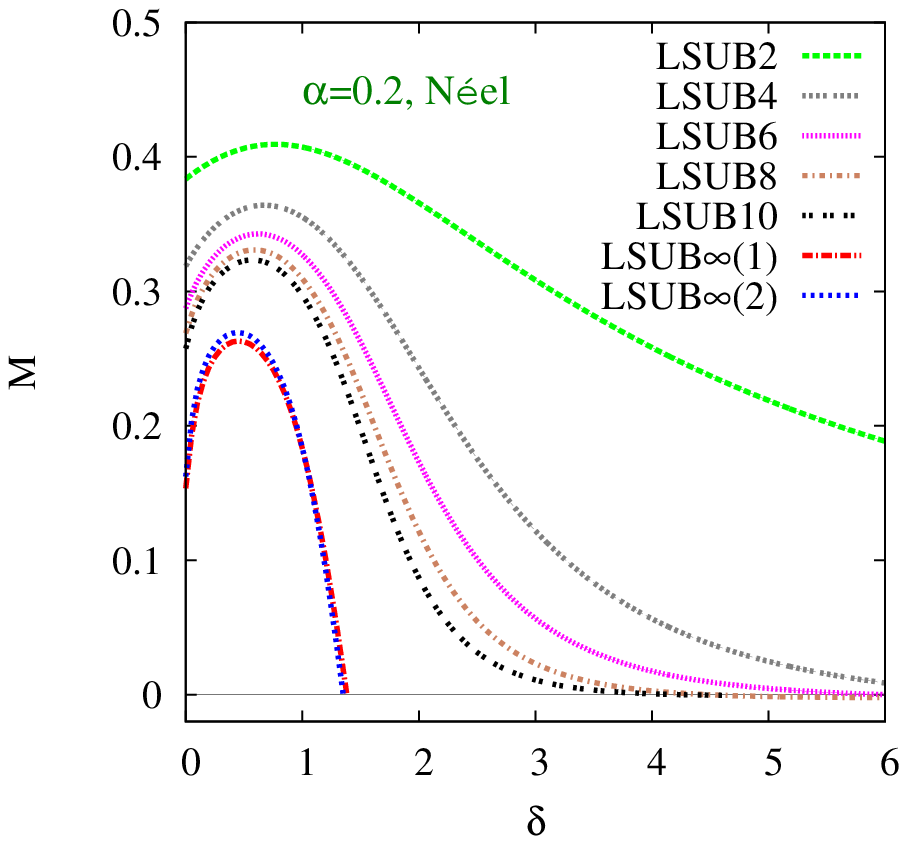}}
\hspace{-1.8cm}
\subfigure[]{\includegraphics[width=7.5cm]{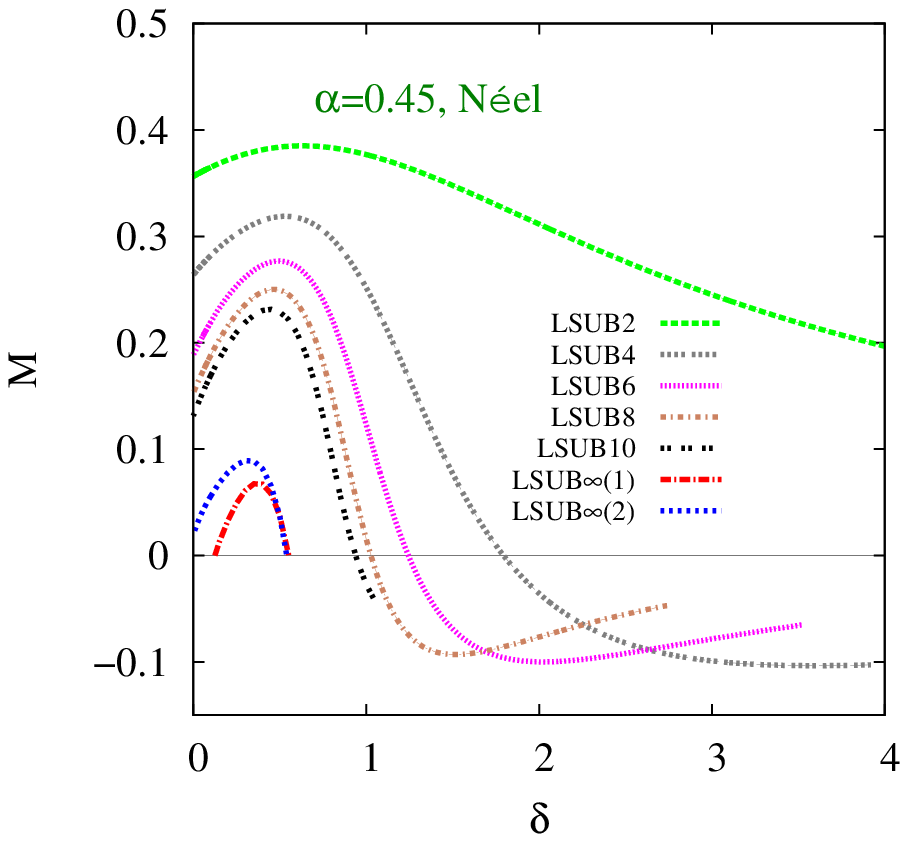}}
}
\caption{CCM results for the GS magnetic order parameter $M$ versus
  the scaled interlayer exchange coupling constant,
  $\delta \equiv J_{1}^{\perp}/J_{1}$, for the spin-$\frac{1}{2}$
  $J_{1}$--$J_{2}$--$J_{3}$--$J_{1}^{\perp}$ model on the bilayer honeycomb
  lattice (with $J_{3}=J_{2}$ and $J_{1}>0$), for three selected values of the
  intralayer frustration parameter, $\alpha \equiv J_{2}/J_{1}$: (a)
  $\alpha=0$, (b) $\alpha=0.2$, and (c) $\alpha=0.45$.  Results based
  on the N\'{e}el state as CCM model state are shown in LSUB$n$
  approximations with $n=2,4,6,8,10$, together with two corresponding
  LSUB$\infty(i)$ extrapolated results using Eq.\
  (\ref{M_extrapo_frustrated}) and the respective data sets
  $n=\{2,6,10\}$ for $i=1$ and $n=\{4,6,8,10\}$ for $i=2$.  In Fig.\ \ref{M_neel_raw_extrapo_fix-J2}(a) we also show, for comparison, an LSUB$\infty(2a)$ extrapolation based on Eq.\ (\ref{M_extrapo_standard}) and the data set $n=\{4,6,8,10\}$.}
\label{M_neel_raw_extrapo_fix-J2}
\end{figure*}

Finally, to obtain estimates for our results in the formally exact
limit, $n \rightarrow \infty$, we need to extrapolate the raw LSUB$n$
data.  Such extrapolations thereby comprise the sole approximation
that we make.  While exact extrapolation rules are not known, much
practical experience from applications of the LSUB$n$ hierarchy to
many different spin-lattice models, has shown the widespread accuracy
of the consistent use of rather simple schemes for the relevant
physical parameters.  In this respect the magnetic order parameter $M$
is of special interest, since two different schemes have been employed
for differing situations.

Thus, for unfrustrated spin systems or for ones with only small amounts of frustration, an appropriate extrapolation scheme is found to be \cite{PHYLi:2012_honeycomb_J1neg,Bishop:2012_honeyJ1-J2,Bishop:2012_honey_circle-phase,RFB:2013_hcomb_SDVBC,DJJFarnell:2014_archimedeanLatt,Bishop:2000_XXZ,Kruger:2000_JJprime,Fa:2001_SqLatt_s1,Darradi:2005_Shastry-Sutherland,Bi:2009_SqTriangle,Bishop:2010_UJack,Bishop:2010_KagomeSq,Bishop:2011_UJack_GrtSpins,Bishop:2017_honeycomb_bilayer_J1J2J1perp}
\begin{equation}
M(n) = m_{0}+m_{1}n^{-1}+m_{2}n^{-2}\,,   \label{M_extrapo_standard}
\end{equation}
from fits to which we obtain the extrapolated LSUB$\infty$ value
$m_{0}$ for $M$.  By contrast, a more appropriate scheme for systems that exhibit a GS order-disorder QPT, or for phases whose order parameter $M$ is zero or small, is \cite{DJJF:2011_honeycomb,PHYLi:2012_honeycomb_J1neg,Bishop:2012_honeyJ1-J2,Bishop:2012_honey_circle-phase,Li:2012_honey_full,RFB:2013_hcomb_SDVBC,Li:2016_honey_grtSpins,Li:2016_honeyJ1-J2_s1,DJJFarnell:2014_archimedeanLatt,Bi:2008_EPL_J1J1primeJ2_s1,Bi:2008_JPCM_J1xxzJ2xxz_s1,Li:2012_anisotropic_kagomeSq,Bishop:2017_honeycomb_bilayer_J1J2J1perp,Li:2018_crossStripe_low-E-param}
\begin{equation}
M(n) = \mu_{0}+\mu_{1}n^{-1/2}+\mu_{2}n^{-3/2}\,,   \label{M_extrapo_frustrated}
\end{equation}
which yields the respective LSUB$\infty$ extrapolant $\mu_{0}$ for
$M$.  

For the triplet spin gap $\Delta$ an extrapolation scheme with
leading power of $n^{-1}$, like that in Eq.\
(\ref{M_extrapo_standard}) for $M$, has been found to give a very good fit
to the LSUB$n$ approximants $\Delta(n)$ for both of the above cases of
unfrustrated (or slightly frustrated) and highly frustrated systems
\cite{Li:2016_honeyJ1-J2_s1,Bishop:2015_honey_low-E-param,Kruger:2000_JJprime,Richter:2015_ccm_J1J2sq_spinGap,Bishop:2015_J1J2-triang_spinGap,Bishop:2017_honeycomb_bilayer_J1J2J1perp,Li:2018_crossStripe_low-E-param},
\begin{equation}
\Delta(n) = d_{0}+d_{1}n^{-1}+d_{2}n^{-2}\,,   \label{Eq_spin_gap}
\end{equation}
from fits to which we obtain the extrapolated LSUB$\infty$ estimate
$d_{0}$ for $\Delta$.

Clearly, for each of the fits of Eqs.\
(\ref{M_extrapo_standard})--(\ref{Eq_spin_gap}), it is best to use
four or more fitting points (i.e., different $n$ values of the LSUB$n$
sequence) to obtain robust and the most reliable results.
Furthermore, the LSUB2 result is expected, {\it a priori} to be too
close to mean-field theory and too far from the exact
$n \rightarrow \infty$ limit to be used in each fit, if it can be
avoided.  For this reason, our preferred sets of LSUB$n$ approximants
for the fits are those with $n=\{4,6,8,10\}$ in the present case where
it is computationally infeasible to calculate results for $n > 10$.

We also note, however, that a $(4m-2)/4m$ staggering effect, where
$m \in \mathbb{Z}^{+}$ is a positive integer, has been observed
\cite{Bishop:2012_honeyJ1-J2,RFB:2013_hcomb_SDVBC,Li:2016_honeyJ1-J2_s1}
in LSUB$n$ sequences of CCM results for various physical parameters on
frustrated honeycomb-lattice monolayers.  Such staggering implies that
the two sub-sequences with $n=4m-2$ and with $n=4m$ need to be
extrapolated separately from one another.  In some cases corresponding
adjacent pairs of curves from each sub-sequence (e.g., those with
$n=2$ and $n=4$, or with $n=6$ and $n=8$) even cross one another as
some coupling parameter is varied.  Such staggering has been possibly
attributed \cite{Li:2016_honeyJ1-J2_s1} to the fact that the honeycomb
lattice comprises two interlocking triangular Bravais lattices, on
each of which a more well-known $(2m-1)/2m$ (i.e., odd/even)
staggering effect is commonly seen, exactly analogous to the same
effect that is well understood in perturbation theory.  Indeed, this
is precisely the reason why LSUB$n$ results with odd values of the
truncation index $n$ are not included in our CCM extrapolations here.
In view of these prior observations on frustrated honeycomb-lattice
monolayers, we shall also compare our results in Sec.\
\ref{results_sec} between extrapolations based on the preferred
sequences with $n=\{4,6,8,10\}$ and those based on $n=\{2,6,10\}$.
The latter sequence is sub-optimal in the two aspects that it both
includes the LSUB2 result and is based on only three fitting points to
extract three parameters.  Nevertheless, it avoids mixing results from
the two staggered sub-sequences.

\section{RESULTS}
\label{results_sec}
We first show in Fig.\ \ref{M_neel_raw_extrapo_fix-J2} our CCM results
for the GS magnetic order parameter $M$ of Eq.\
(\ref{M_definition_eq}) in the N\'{e}el phase, as a function of the
scaled interlayer exchange coupling constant,
$\delta \equiv J_{1}^{\perp}/J_{1}$, for three particular
representative values of the intralayer frustration parameter,
$\alpha \equiv J_{2}/J_{1}$.
In each case we show the ``raw'' LSUB$n$ data with $n=2,4,6,8,10$,
based on the N\'{e}el state of Fig.\ \ref{model_pattern}(c) as the CCM
model state, together with various LSUB$\infty$ extrapolations.  We
note in particular that the two extrapolations based on the scheme of
Eq.\ (\ref{M_extrapo_frustrated}), but using the two respective
LSUB$n$ data sets with $n=\{2,6,10\}$ and $n=\{4,6,8,10\}$ give
results in very close agreement with one another for all three values
of $\alpha$ shown and for most values of $\delta$.  Generally, the
only exception, where there is some slight sensitivity to the
extrapolation input data is the joint region of the highest values of
intralayer frustration (near to where N\'{e}el order disappears) and
the lowest values of interlayer AFM coupling, as seen in Fig.\
\ref{M_neel_raw_extrapo_fix-J2}(c), for example.

\begin{figure*}[t]
\mbox{
\hspace{-1.0cm}
\subfigure[]{\includegraphics[width=7.5cm]{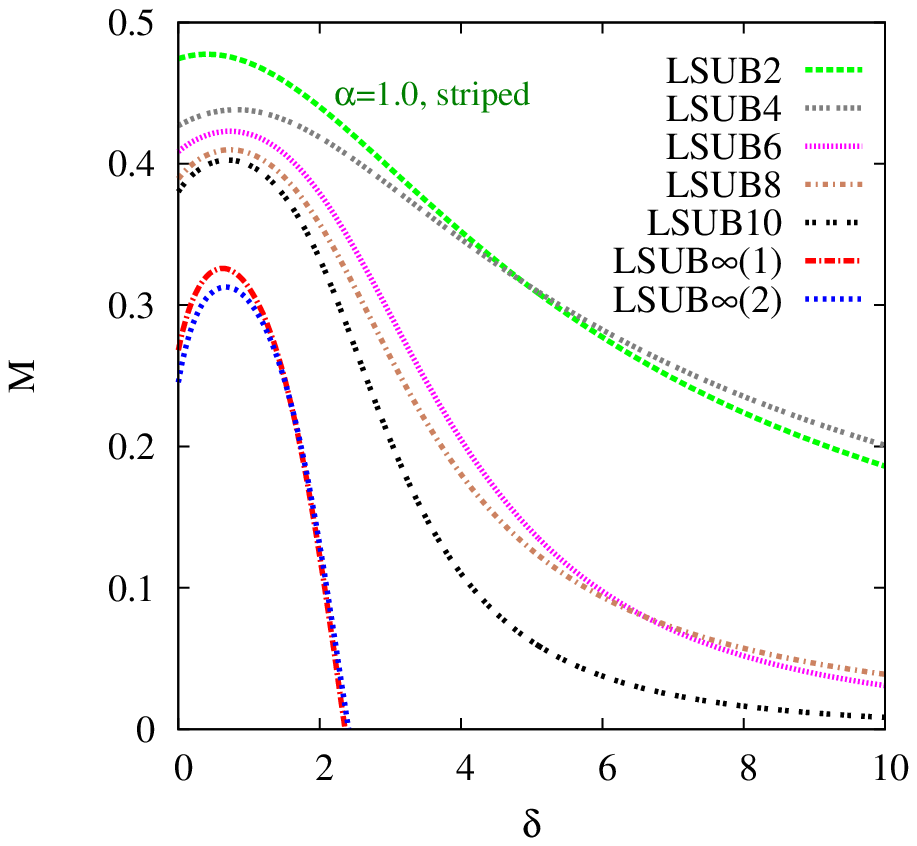}}
\hspace{-1.8cm}
\subfigure[]{\includegraphics[width=7.5cm]{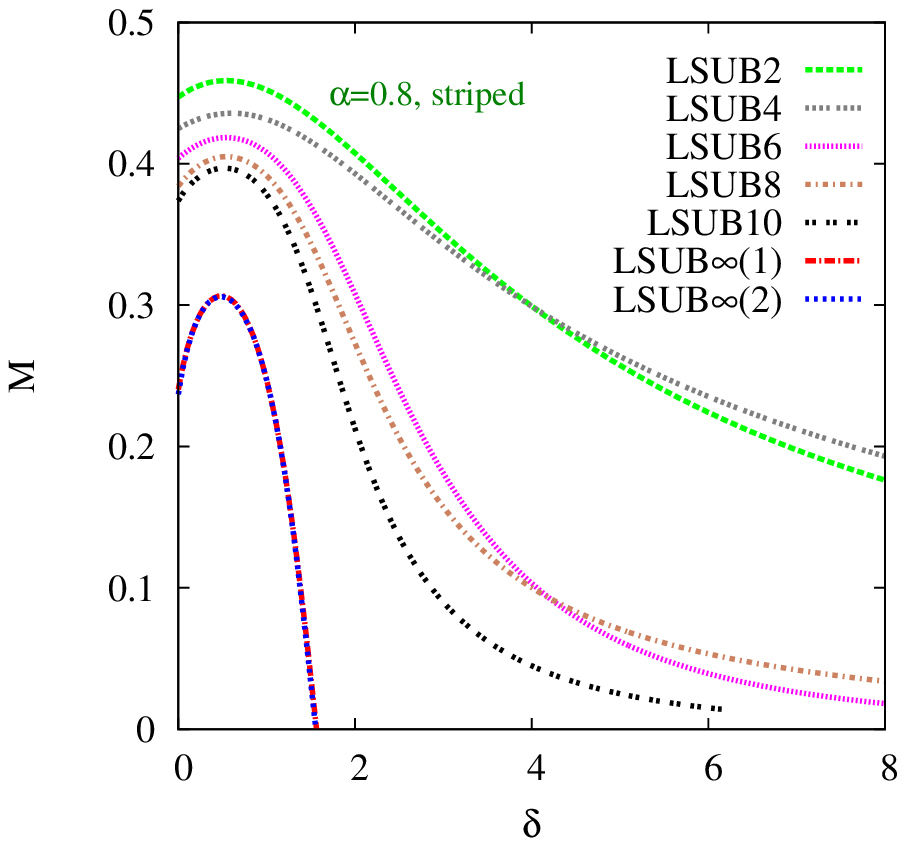}}
\hspace{-1.8cm}
\subfigure[]{\includegraphics[width=7.5cm]{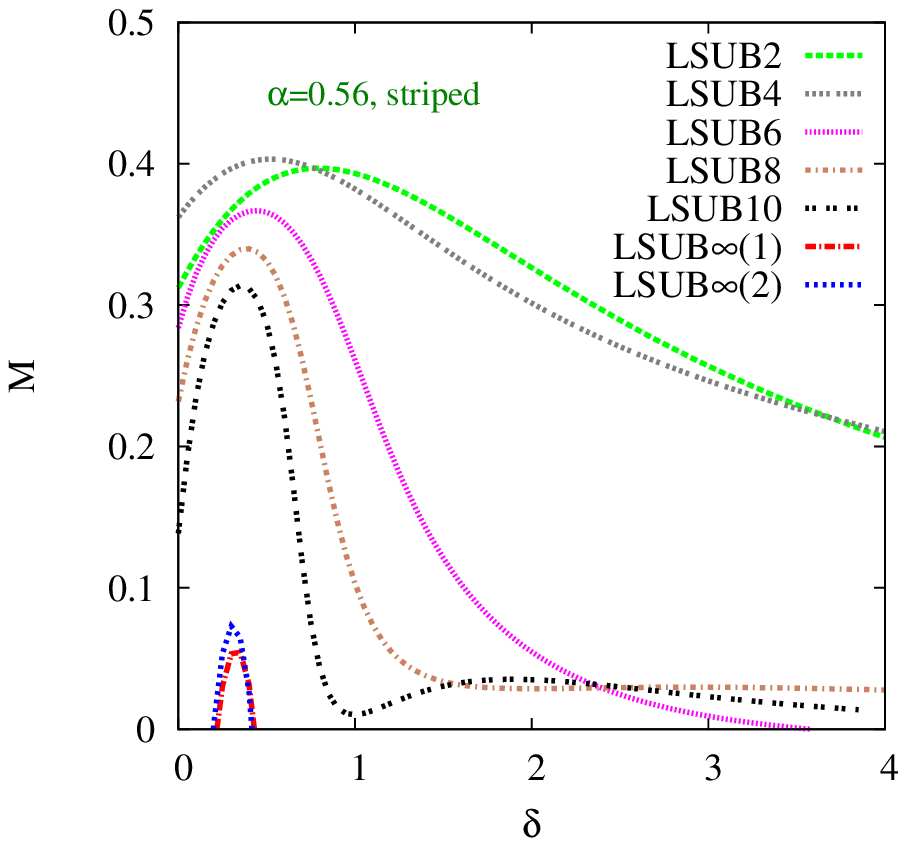}}
}
\caption{CCM results for the GS magnetic order parameter $M$ versus
  the scaled interlayer exchange coupling constant,
  $\delta \equiv J_{1}^{\perp}/J_{1}$, for the spin-$\frac{1}{2}$
  $J_{1}$--$J_{2}$--$J_{3}$--$J_{1}^{\perp}$ model on the bilayer honeycomb
  lattice (with $J_{3}=J_{2}$ and $J_{1}>0$), for three selected values of the
  intralayer frustration parameter, $\alpha \equiv J_{2}/J_{1}$: (a)
  $\alpha=1.0$, (b) $\alpha=0.8$, and (c) $\alpha=0.56$.  Results based
  on the striped state as CCM model state are shown in LSUB$n$
  approximations with $n=2,4,6,8,10$, together with two corresponding
  LSUB$\infty(i)$ extrapolated results using Eq.\
  (\ref{M_extrapo_frustrated}) and the respective data sets
  $n=\{2,6,10\}$ for $i=1$ and $n=\{4,6,8,10\}$ for $i=2$.}
\label{M_striped_raw_extrapo_fix-J2}
\end{figure*}

It is interesting to note that in each case the effect of turning on
the interlayer coupling is first to enhance the stability of the
N\'{e}el order, up to some particular value of $\delta$, which depends
on the intralayer frustration parameter $\alpha$.  Increasing $\delta$
beyond this value then leads to a decrease in the N\'{e}el order
parameter $M$, out to some critical value $\delta_{c_{1}}^{>}(\alpha)$
at which $M \rightarrow 0$.  Furthermore, we note that N\'{e}el order
persists in the honeycomb-lattice monolayer $(\delta=0)$ for all
values of the frustration parameter in the range
$\alpha < \alpha_{c_{1}}(0)$.  In this range we find an upper critical
value $\delta_{c_{1}}^{>}(\alpha)$ of the scaled interlayer exchange
coupling constant, such that N\'{e}el order persists over the range
$0<\delta<\delta_{c_{1}}^{>}(\alpha)$.  Very interestingly, however,
as may be seen from Fig.\ \ref{M_neel_raw_extrapo_fix-J2}(c), for
example, for higher values of $\alpha$ in the range
$\alpha_{c_{1}}(0) < \alpha < \alpha_{1}^{>}$, we find a reentrant
type of behavior in which N\'{e}el order now exists only over the
range
$\delta_{c_{1}}^{<}(\alpha) < \delta < \delta_{c_{1}}^{>}(\alpha)$,
with $\delta_{c_{1}}^{<}(\alpha) > 0$.  The corresponding upper and lower
critical values coincide when $\alpha = \alpha_{1}^{>}$, at which
point, we thus have
$\delta_{c_{1}}^{<}(\alpha_{1}^{>})=\delta_{c_{1}}^{>}(\alpha_{1}^{>})$.
Finally, N\'{e}el order is absent for $\alpha > \alpha_{1}^{>}$, for
all values of $\delta$.

The extrapolation scheme of Eq.\ (\ref{M_extrapo_frustrated}), used
for the LSUB$\infty(1)$ and LSUB$\infty(2)$ results shown in each of
Figs.\ \ref{M_neel_raw_extrapo_fix-J2}(a),
\ref{M_neel_raw_extrapo_fix-J2}(b), and
\ref{M_neel_raw_extrapo_fix-J2}(c), is certainly valid for all these
cases when the frustration is appreciable, especially for systems
close to a QCP as here, at which the magnetic order parameter
vanishes.  However, for unfrustrated systems or ones that are only
slightly unfrustrated, the scheme of Eq.\ (\ref{M_extrapo_standard})
provides a better fit to the LSUB$n$ input data.  Accordingly, we also
show in Fig.\ \ref{M_neel_raw_extrapo_fix-J2}(a) alone the
LSUB$\infty(2a)$ extrapolation based on Eq.\
(\ref{M_extrapo_standard}) and the input data set $n=\{4,6,8,10\}$.
This extrapolation is then valid for this case of zero intralayer
frustration $(\alpha=0)$ only for a small range of values of the
scaled interlayer coupling, $\delta \lesssim 0.2$, say.

In particular, the LSUB$\infty(2a)$ curve gives a value $M=0.2741(1)$
for the unfrustrated honeycomb-lattice monolayer (i.e., with
$\alpha=0=\delta$), where the quoted error is simply the least-squares
error associated with the fit.  The corresponding result using Eq.\
(\ref{M_extrapo_standard}) and the input data set $n=\{2,6,10\}$ is
$M=0.2761$.  There is clearly a small sensitivity associated with the
LSUB$n$ input data set used of about $1\%$ or so.  In this case
$(\alpha=0=\delta)$ alone, where ``minus-sign problems'' are absent,
we may compare our CCM results with recent values obtained from two
different quantum Monte Carlo (QMC) simulations, which yielded the
corresponding respective values $M=0.2681(8)$ \cite{Low:2009_honey}
and $M=0.26882(3)$ \cite{Jiang:2012_honey}.  The agreement with our
own CCM estimates is very good, and we expect a corresponding accuracy
(i.e., of the order of $2\%$) for all the results that we present.
Indeed, for the monolayer case, where we are also able to perform
LSUB$n$ calculations with $n=12$, our corresponding extrapolated
value, based on the data set $n=\{8,10,12\}$, again with only three
fitting points, for example, is $M=0.2715$, in even better agreement
$(1\%)$ with the QMC result.

In Fig.\ \ref{M_striped_raw_extrapo_fix-J2} we show corresponding
results for the GS magnetic order parameter $M$ of Eq.\ 
(\ref{M_definition_eq}) in the striped phase to those shown in Fig.\
\ref{M_neel_raw_extrapo_fix-J2} for the N\'{e}el phase.
Again, we show results for $M(\delta)$ for three particular
representative values of the intralayer frustration parameter
$\alpha$.  In each case we show the same ``raw'' LSUB$n$ data with
$n=2,4,6,8,10$, but now based on the striped state of Fig.\
\ref{model_pattern}(d) as the CCM model state.  We also display in
Fig.\ \ref{M_striped_raw_extrapo_fix-J2} the same LSUB$\infty(1)$ and
LSUB$\infty(2)$ extrapolations as those shown in Fig.\
\ref{M_neel_raw_extrapo_fix-J2}, both of which are based on the
appropriate scheme of Eq.\ (\ref{M_extrapo_frustrated}), but with the
two respective LSUB$n$ input data sets $n=\{2,6,10\}$ and
$n=\{4,6,8,10\}$.

Again, we note that the two extrapolations are in remarkable agreement
with each other in every case, despite the $(4m-2)/4m$ staggering that
has been discussed in Sec.\ \ref{ccm_sec} and which is now clearly
visible in each of Figs.\ \ref{M_striped_raw_extrapo_fix-J2}(a),
\ref{M_striped_raw_extrapo_fix-J2}(b), and
\ref{M_striped_raw_extrapo_fix-J2}(c), where it manifests itself most
vividly as actual crossings of corresponding adjacent pairs of curves
(viz., LSUB2 and LSUB4, corresponding to $m=1$, and LSUB6 and LSUB8,
corresponding to $m=2$).  Presumably, the very close agreement between
the LSUB$\infty(1)$ and LSUB$\infty(2)$ extrapolations, the former of
which takes the staggering into account and the latter of which does
not, is related to the fact that the crossings occur only for
unphysical values of $\delta$, far beyond the associated QCP at which
the (extrapolated) striped order parameter has vanished.

\begin{figure*}[t]
\mbox{
\subfigure[]{\includegraphics[width=8.5cm]{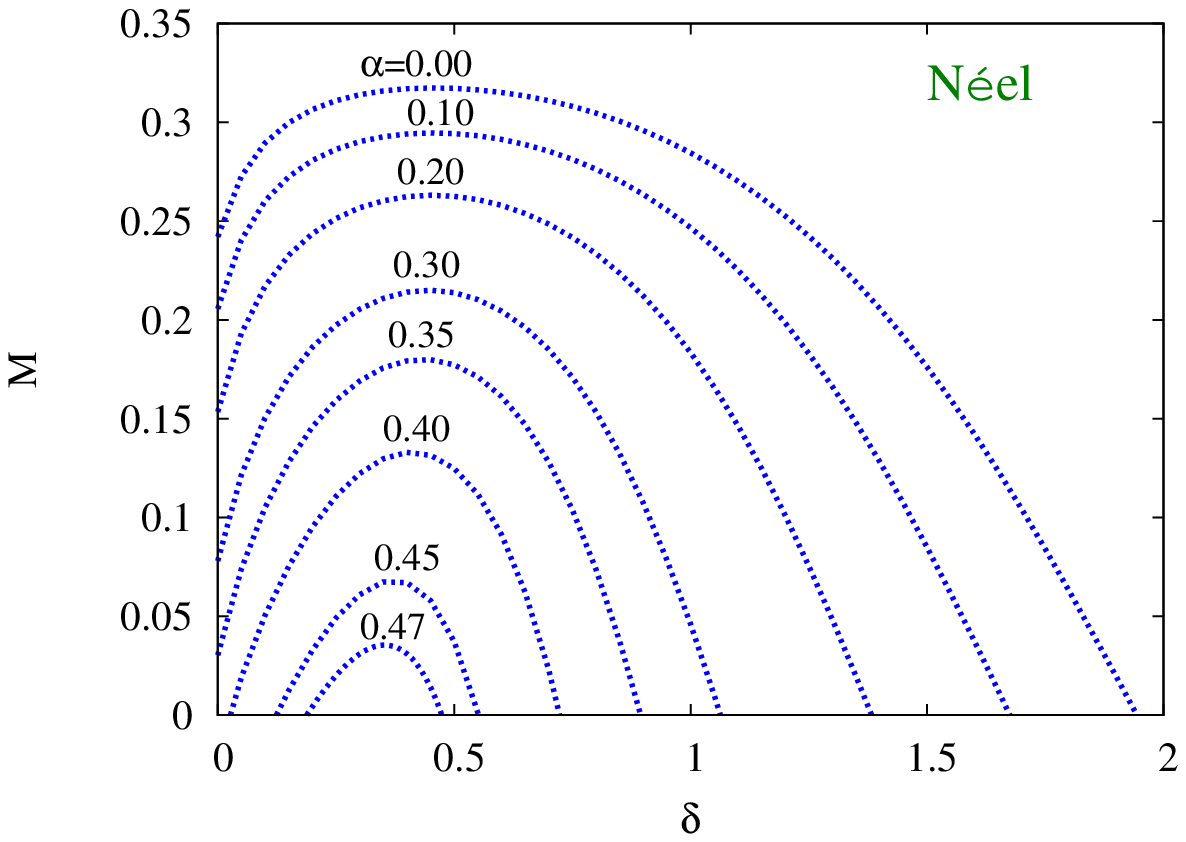}}
\subfigure[]{\includegraphics[width=8.5cm]{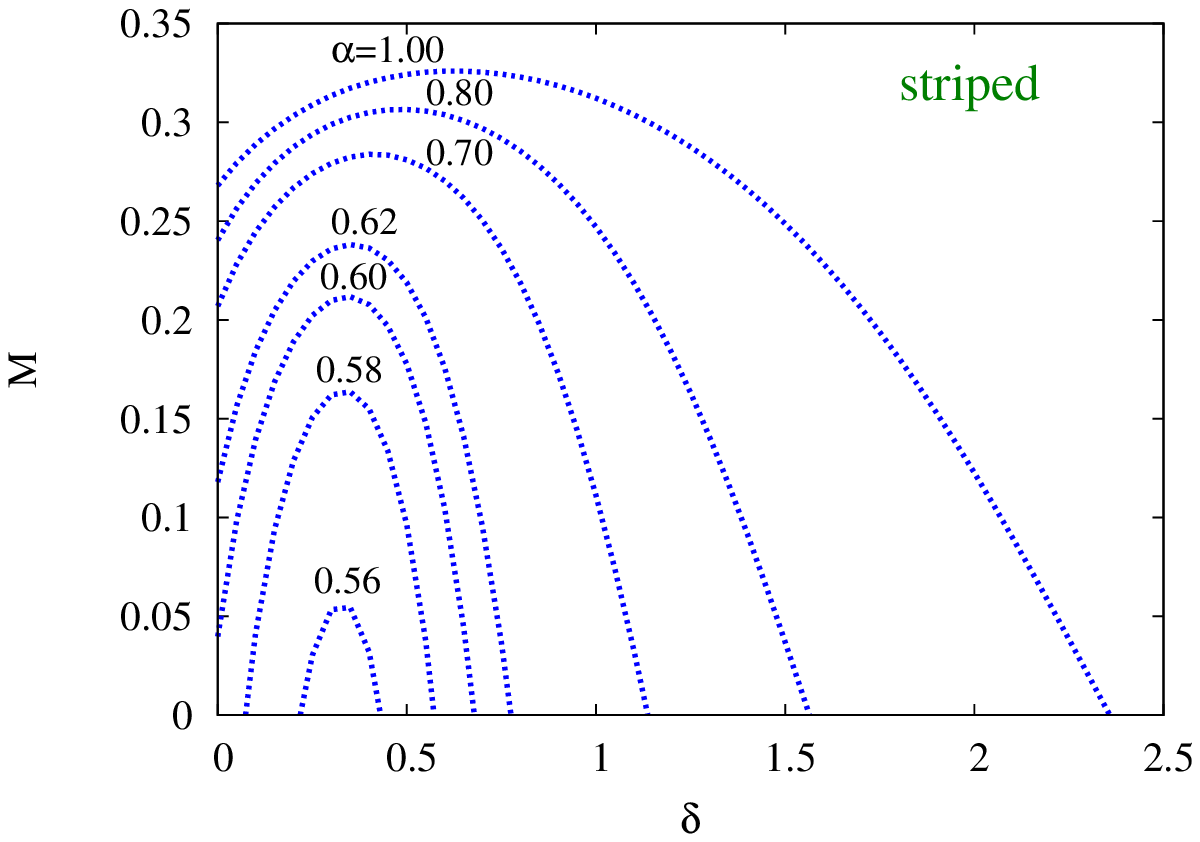}}
  }
  \caption{CCM results for the GS magnetic order parameter $M$ versus
    the scaled interlayer exchange coupling constant,
    $\delta \equiv J_{1}^{\perp}/J_{1}$, for the spin-$\frac{1}{2}$
    $J_{1}$--$J_{2}$--$J_{3}$--$J_{1}^{\perp}$ model on the bilayer honeycomb
    lattice (with $J_{3}=J_{2}$ and $J_{1}>0$), for a variety of values of the
    intralayer frustration parameter, $\alpha \equiv J_{2}/J_{1}$, using (a) the N\'{e}el state and (b) the striped state as the CCM model state.
  In each case we show extrapolated results, obtained from using Eq.\
  (\ref{M_extrapo_frustrated}) with the corresponding LSUB$n$ data
  sets $n=\{2,6,10\}$.}
\label{M_J2fix-selective_extrapo}
\end{figure*}

Figure \ref{M_striped_raw_extrapo_fix-J2} exhibits a reentrant
behavior very similar to that seen in Fig.\
\ref{M_neel_raw_extrapo_fix-J2}, and discussed above, for the N\'{e}el
phase.  Thus, striped order is present in the honeycomb-lattice
monolayer $(\delta=0)$ for all values of the intralayer frustration
parameter in the range $\alpha > \alpha_{c_{2}}(0)$.  In this range we
observe from Fig.\ \ref{M_striped_raw_extrapo_fix-J2} that as the AFM
bilayer coupling is turned on, striped order persists over the range
$0 < \delta < \delta_{c_{2}}^{>}(\alpha)$ of the scaled interlayer
exchange coupling.  However, now for somewhat lower values of $\alpha$
in the range $\alpha_{2}^{<} < \alpha <\alpha_{c_{2}}(0)$, we again
observe a reentrant behavior in which striped order reappears over the
range
$\delta_{c_{2}}^{<}(\alpha) < \delta < \delta_{c_{2}}^{>}(\alpha)$,
with $\delta_{c_{2}}^{<}(\alpha) > 0$.  Similar to the N\'{e}el case,
these respective upper and lower critical values for the striped phase
coalesce when $\alpha = \alpha_{2}^{<}$, such that
$\delta_{c_{2}}^{<}(\alpha_{2}^{<}) =
\delta_{c_{2}}^{>}(\alpha_{2}^{<})$.  Striped order is then absent for
all values of $\delta$ for $\alpha < \alpha_{2}^{<}$.

The reentrant behavior for both the N\'{e}el and striped phases is
also demonstrated more graphically in Fig.\
\ref{M_J2fix-selective_extrapo}.  Here we show sequences of
extrapolated results for the magnetic order parameter of both phases
as functions of $\delta$, for a variety of values of $\alpha$ in both
cases.  All of the extrapolations shown are based on the scheme of
Eq.\ (\ref{M_extrapo_frustrated}), used with the corresponding LSUB$n$
input data sets with $n=\{2,6,10\}$.  With this particular
extrapolation the two QCPs for the monolayers are
$\alpha_{c_{1}}(0) \approx 0.379$ and
$\alpha_{c_{2}}(0) \approx 0.595$.  These may be compared, for
example, with the corresponding results \cite{Cabra:2011_honey},
$\alpha_{c_{1}}(0) \approx 0.41$ and $\alpha_{c_{2}}(0) \approx 0.6$,
from using a rotationally-invariant version of Schwinger boson
mean-field theory, which has proven itself to be a fairly accurate
technique for taking quantum fluctuations into account.  By contrast,
lowest-order (or linear) spin-wave theory gives the less accurate
results \cite{Cabra:2011_honey}, $\alpha_{c_{1}}(0) \approx 0.29$ and
$\alpha_{c_{2}}(0) \approx 0.55$.  Our own corresponding CCM results
for $\alpha_{1}^{>}$ and $\alpha_{2}^{<}$ are
$\alpha_{1}^{>} \approx 0.487$ and $\alpha_{2}^{<} \approx 0.556$,
again based on the extrapolation scheme of Eq.\
(\ref{M_extrapo_frustrated}) used with the LSUB$n$ input data set with
$n=\{2,6,10\}$.

\begin{figure*}[t]
\mbox{
\hspace{-1.0cm}
\subfigure[]{\includegraphics[width=7.5cm]{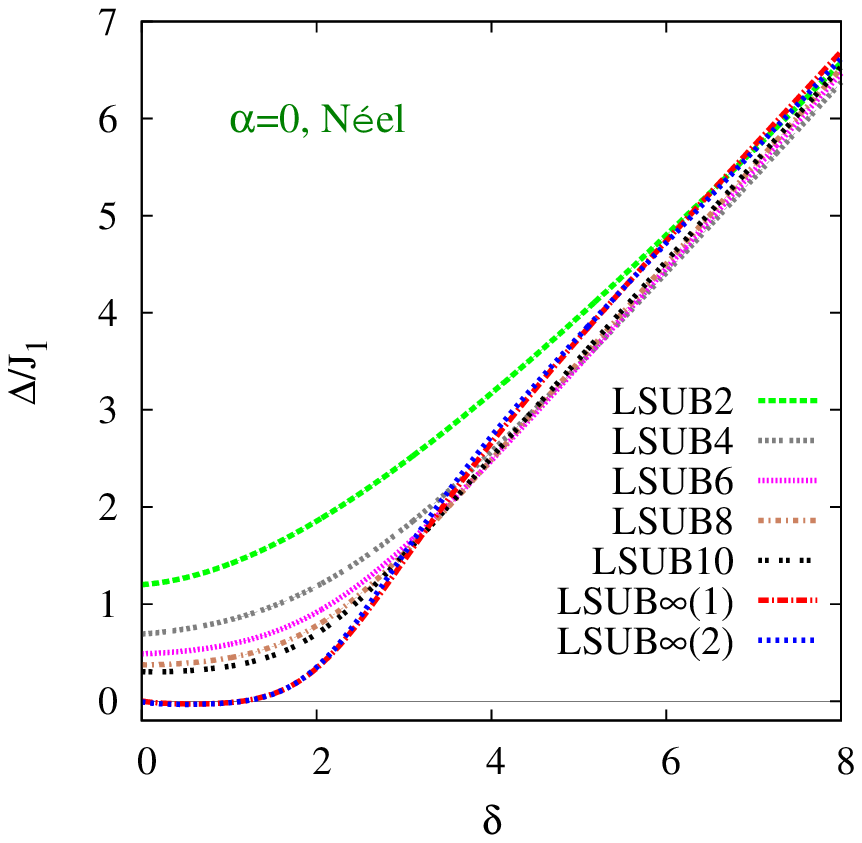}}
\hspace{-1.8cm}
\subfigure[]{\includegraphics[width=7.5cm]{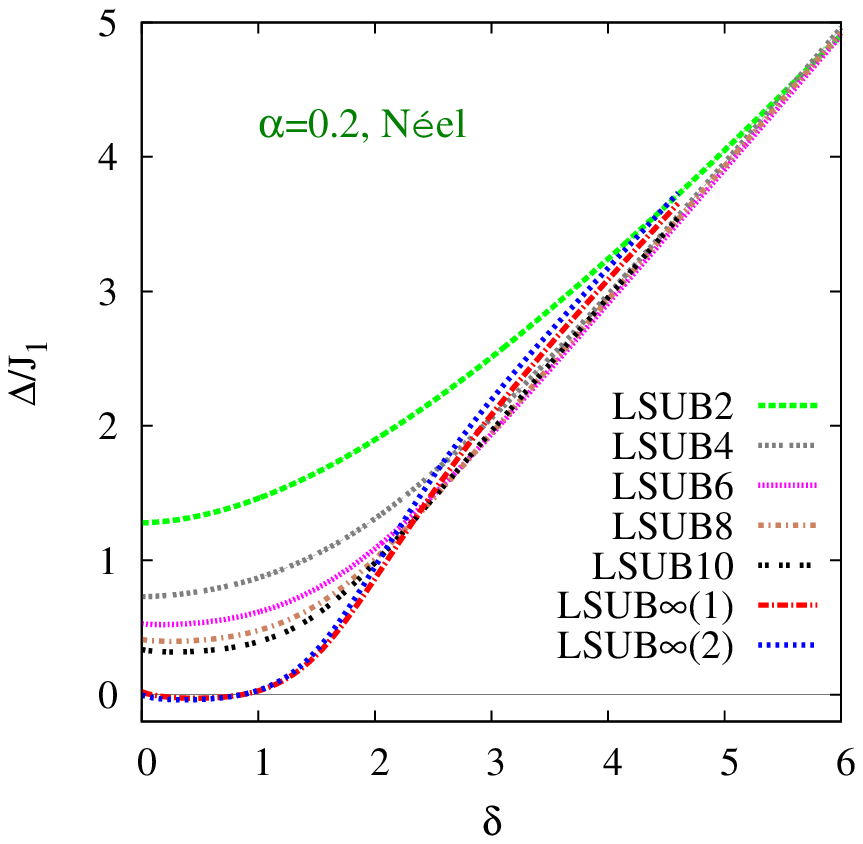}}
\hspace{-1.8cm}
\subfigure[]{\includegraphics[width=7.5cm]{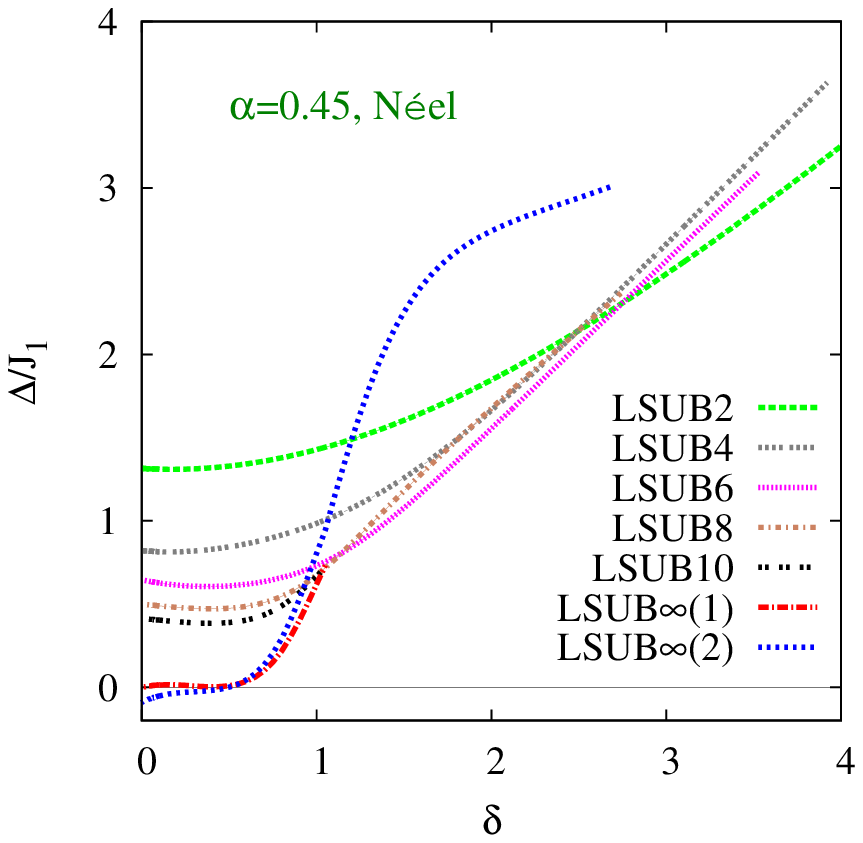}}
}
\caption{CCM results for the triplet spin gap $\Delta$ (in units of
  $J_{1}$) versus the scaled interlayer exchange coupling constant,
  $\delta \equiv J_{1}^{\perp}/J_{1}$, for the spin-$\frac{1}{2}$
  $J_{1}$--$J_{2}$--$J_{3}$--$J_{1}^{\perp}$ model on the bilayer
  honeycomb lattice (with $J_{3}=J_{2}$ and $J_{1}>0$), for three
  selected values of the intralayer frustration parameter,
  $\alpha \equiv J_{2}/J_{1}$: (a) $\alpha=0$, (b) $\alpha=0.2$, and
  (c) $\alpha=0.45$.  Results based on the N\'{e}el state as CCM model
  state are shown in LSUB$n$ approximations with $n=2,4,6,8,10$,
  together with two corresponding LSUB$\infty(i)$ extrapolated results
  using Eq.\ (\ref{Eq_spin_gap}) and the respective data sets
  $n=\{2,6,10\}$ for $i=1$ and $n=\{4,6,8,10\}$ for $i=2$.}
\label{Egap_neel_raw_extrapo_fix-J2}
\end{figure*}
\begin{figure*}
\mbox{
\hspace{-1.0cm}
\subfigure[]{\includegraphics[width=7.5cm]{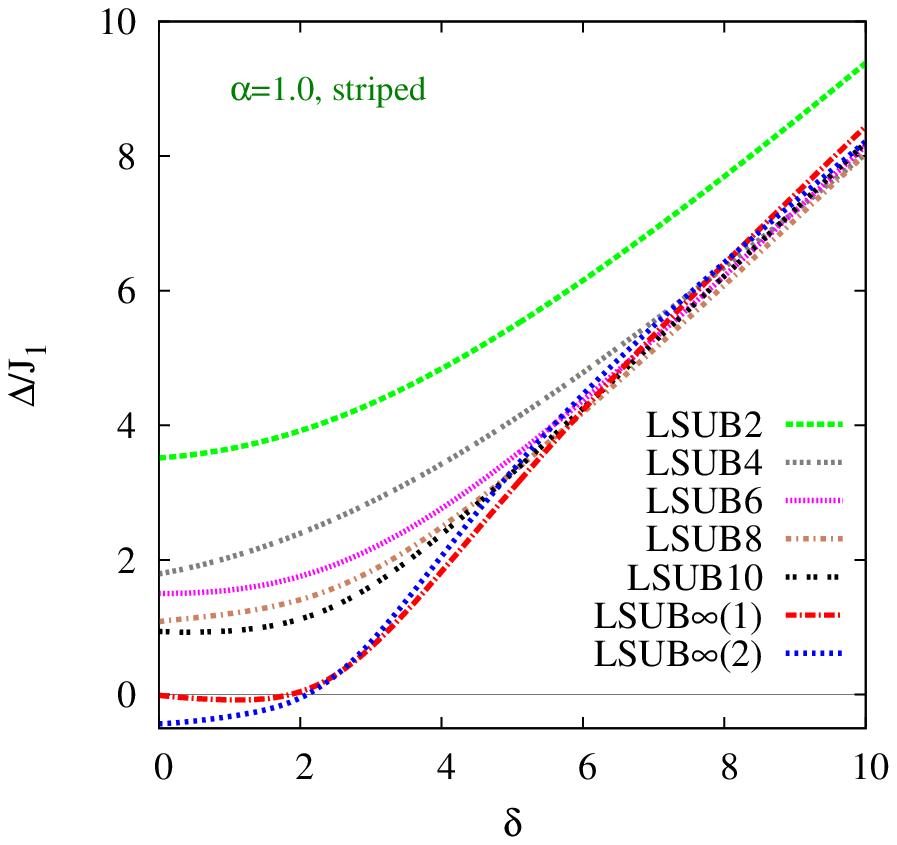}}
\hspace{-1.8cm}
\subfigure[]{\includegraphics[width=7.5cm]{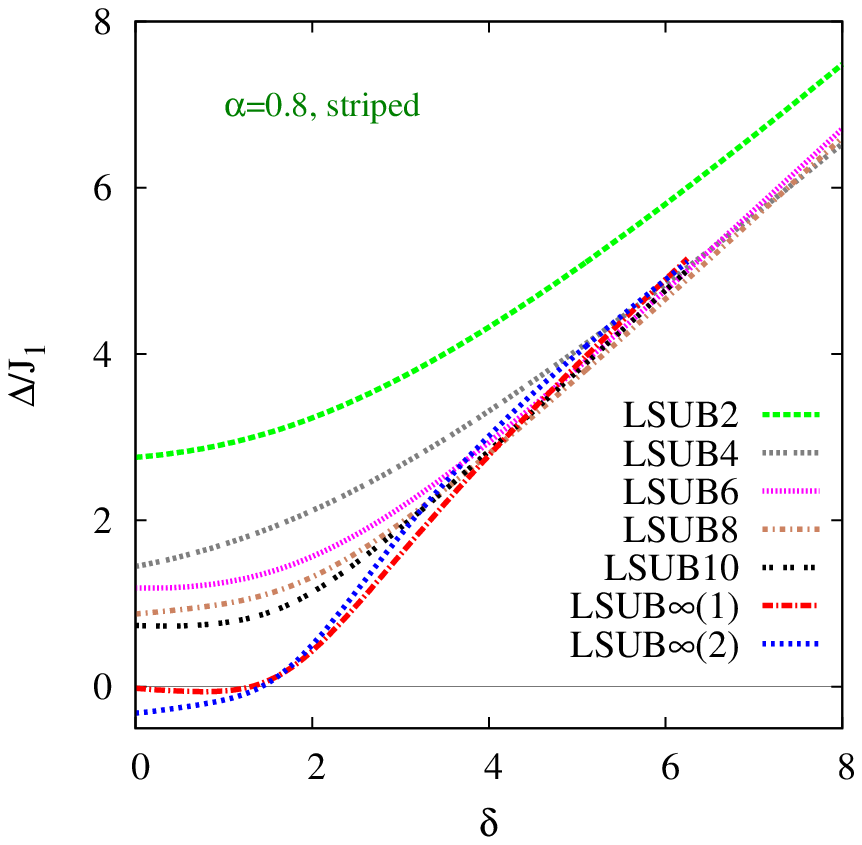}}
\hspace{-1.8cm}
\subfigure[]{\includegraphics[width=7.5cm]{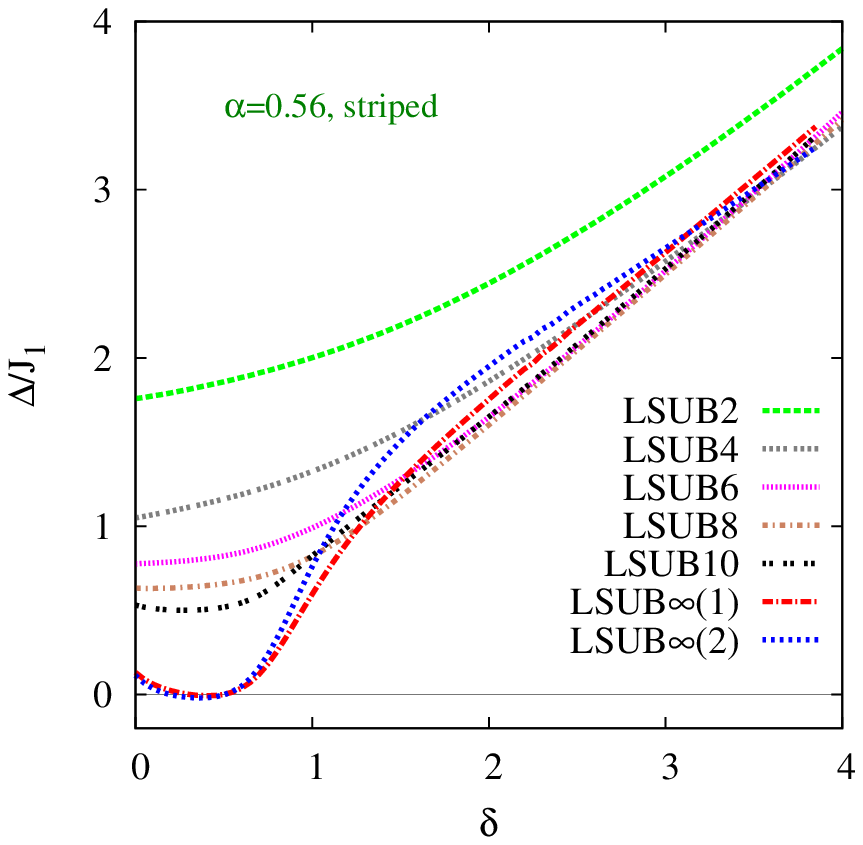}}
}
\caption{CCM results for the triplet spin gap $\Delta$ (in units of $J_{1}$) versus
  the scaled interlayer exchange coupling constant,
  $\delta \equiv J_{1}^{\perp}/J_{1}$, for the spin-$\frac{1}{2}$
  $J_{1}$--$J_{2}$--$J_{3}$--$J_{1}^{\perp}$ model on the bilayer honeycomb
  lattice (with $J_{3}=J_{2}$ and $J_{1}>0$), for three selected values of the
  intralayer frustration parameter, $\alpha \equiv J_{2}/J_{1}$: (a)
  $\alpha=1.0$, (b) $\alpha=0.8$, and (c) $\alpha=0.56$.  Results based
  on the striped state as CCM model state are shown in LSUB$n$
  approximations with $n=2,4,6,8,10$, together with two corresponding
  LSUB$\infty(i)$ extrapolated results using Eq.\
  (\ref{Eq_spin_gap}) and the respective data sets
  $n=\{2,6,10\}$ for $i=1$ and $n=\{4,6,8,10\}$ for $i=2$.}
\label{Egap_striped_raw_extrapo_fix-J2}
\end{figure*}

We turn now to our respective results for the triplet spin gap
$\Delta$.  We first show in Fig.\ \ref{Egap_neel_raw_extrapo_fix-J2}
results based on the N\'{e}el state as our CCM model state as a
function of the scaled interlayer exchange coupling constant,
$\delta \equiv J_{1}^{\perp}/J_{1}$, for the same three representative
values of the intralayer frustration parameter,
$\alpha \equiv J_{2}/J_{1}$, as were shown in Fig.\
\ref{M_neel_raw_extrapo_fix-J2} for the GS magnetic order parameter
$M$.  We note that the two sets of extrapolations, now both based on Eq.\
(\ref{Eq_spin_gap}) but using the two LSUB$n$ input data sets with
$n=\{2,6,10\}$ and $n=\{4,6,8,10\}$, are in excellent agreement with
one another.  Both give results for $\Delta$ that are zero, within
very small numerical errors, when N\'{e}el order is present, as
expected, i.e., for $\delta < \delta_{c_{1}}^{>}(\alpha)$.
Furthermore, the values obtained for $\delta_{c_{1}}^{>}(\alpha)$ are
in good agreement with the independent values obtained from Fig.\
\ref{M_neel_raw_extrapo_fix-J2} for where the GS magnetic order
parameter $M$ vanishes.  It is clear too that in each case the
non-magnetic phase that opens up after the melting of N\'{e}el order
is gapped, consistent with it having a VBC character.

Our corresponding CCM results for the triplet spin gap $\Delta$ based
on the striped state as the model state are shown in Fig.\
\ref{Egap_striped_raw_extrapo_fix-J2}.  Once again we display results
as a function of $\delta$, now for the same three representative
values of $\alpha$ as were shown in Fig.\
\ref{M_striped_raw_extrapo_fix-J2} for the magnetic order parameter
$M$.  In this case the extrapolated LSUB$\infty(1)$ results, based on
the input LSUB$n$ data set $n=\{2,6,10\}$, give results that $\Delta$
is zero, within extremely small numerical errors, for all three values
of $\alpha$ shown, over essentially the same ranges of values of $\delta$, viz.,
$0 < \delta < \delta_{c_{2}}^{>}(\alpha)$ for $\alpha > \alpha_{c_{2}}(0)$
and $\delta_{c_{2}}^{<} < \delta < \delta_{c_{2}}^{>}(\alpha)$ for
$\alpha_{2}^{<} < \alpha < \alpha_{c_{2}}(0)$, for which the striped
magnetic order parameter $M$ is positive in Fig.\
\ref{M_striped_raw_extrapo_fix-J2}.  By contrast, for the striped
state, the LSUB$\infty(2)$ extrapolation, based on the input LSUB$n$
data set with $n=\{4,6,8,10\}$, is definitely not as accurate or
reliable as the LSUB$\infty(1)$ extrapolation in this respect.  This
is surely due now to the $(4m-2)/4m$ staggering effect that is clearly
visible in the LSUB$n$ sequences of results shown in Fig.\
\ref{Egap_striped_raw_extrapo_fix-J2}(a)--(c).

By way of further elucidation of the extrapolation of sequences of
approximants that display a staggering as in Fig.\
\ref{Egap_striped_raw_extrapo_fix-J2}, it is instructive to consider
an analogous situation in $n$th-order perturbation theory (PT), for
which exact extrapolation laws can be derived for various physical
quantities.  However, as is very well known, the even ($n=2m$, where
$m \in \mathbb{Z}^{+}$) and odd ($n=2m-1$) sequences of PT
approximants involve an additional staggering effect, exactly as is
also seen in corresponding CCM LSUB$n$ approximants.  In both cases
the even and odd sequences obey an extrapolation scheme of the same
sort (i.e., with the same leading exponent), but one should not mix
even and odd terms together in a single approximation scheme, unless
the staggering is incorporated somehow, since the coefficients are not
identical for both sequences.  The explicit inclusion of the
staggering is always very difficult to achieve in a robust manner, and
hence in practice one always extrapolates only the even-order terms or
the odd-order terms, as mentioned in Sec.\ \ref{ccm_sec}.  For
honeycomb-lattice models of the sort considered here, it has been
observed previously, as discussed above, that there is an {\it
  additional} staggering in the even-order sequence of LSUB$n$ terms
between those with $n=4m$ and those with $n=4m-2$.  This staggering
can clearly be seen by visual inspection of the LSUB$n$ curves shown
in Fig.\ \ref{Egap_striped_raw_extrapo_fix-J2}.  Once again, both
subsequences still separately obey Eq.\ (\ref{Eq_spin_gap}).  If we do,
however, despite the staggering, mix terms from the latter two
subsequences, as in our LSUB$\infty(2)$ extrapolation scheme, we get a
poorer fit, leading, for example, to the observed negative values for
the spin gap in Fig.\ \ref{Egap_striped_raw_extrapo_fix-J2}.  Such an
obviously incorrect result is simply due to the staggering effect not
having been incorporated.

We note, with regard to our CCM spin gap results, that the LSUB$n$
curves in both Figs.\ \ref{Egap_neel_raw_extrapo_fix-J2} and
\ref{Egap_striped_raw_extrapo_fix-J2} clearly show a linear increase
with $\delta$ for large values of this parameter.  This is precisely
as expected for an IDVBC phase, as expressed by Eq.\
(\ref{triplet_spin_gap_scaled}).  We note too that for true quantum
critical behavior we expect the spin gap $\Delta$ to vanish (for a
fixed value of the frustration parameter $\alpha$, say) at some
critical value $\delta_{c}$ of the interlayer coupling as
$\Delta \rightarrow \kappa|\delta-\delta_{c}|^{\varepsilon}$ as
$\delta \rightarrow \delta_{c}$, with a critical exponent
$\varepsilon$ and with $\kappa$ a constant.  Our LSUB$n$ curves shown
in Figs.\ \ref{Egap_neel_raw_extrapo_fix-J2} and
\ref{Egap_striped_raw_extrapo_fix-J2}, both for the ``raw'' curves
with $n$ finite and the extrapolated values with
$n \rightarrow \infty$, are clearly more consistent with a value
$\varepsilon > 1$ (i.e., so that $\Delta$ vanishes with a zero slope,
rather than the infinite slope expected for $\varepsilon < 1$) at both
critical points $\delta_{c_{1}}(\alpha)$ and $\delta_{c_{2}}(\alpha)$.

While the phase boundaries obtained from our CCM results for $M$ and
$\Delta$ for both quasiclassical magnetic phases are thus clearly in
excellent agreement with each other, those obtained from $M$ are
surely more accurate.  This is simply due to the respective shapes of
the curves for $M$ and $\Delta$ as functions of the parameters
$\alpha$ and $\delta$.  Thus, the (extrapolated) curves for $\Delta$,
which are zero (within numerical errors) in the magnetically ordered
phase, generally depart from zero (to indicate a gapped state) with
zero slope.  Thus, the estimates for the QCPs from the results for
$\Delta$ have much larger associated errors than those obtained from
the vanishing of the order parameter $M$, since the slope of the curve
for $M$ as a function of the corresponding coupling constant is
generally nonzero at the points where $M \rightarrow 0$.
\begin{figure}[!t]
  \includegraphics[width=9cm]{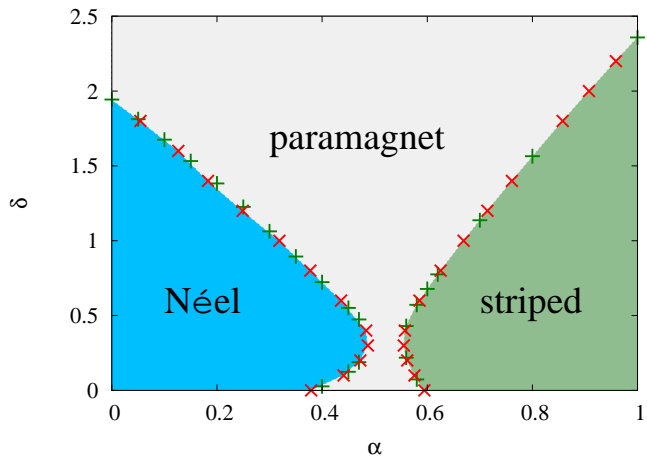}
  \caption{$T=0$ phase diagram of the spin-$\frac{1}{2}$
    $J_{1}$--$J_{2}$--$J_{3}$--$J_{1}^{\perp}$ model on the
    $AA$-stacked bilayer honeycomb lattice with $J_{1}>0$,
    $J_{3}=J_{2}\equiv\alpha J_{1}>0$,
    $J_{1}^{\perp}\equiv\delta J_{1}>0$.  The leftmost (skyblue) and
    the rightmost (seagreen) regions are the quasiclassical AFM phases
    with N\'{e}el and striped order respectively, while the central
    (grey) region is a gapped paramagnetic phase.  The red cross
    ($\times$) and the green plus ($+$) symbols are points at which
    the extrapolated GS magnetic order parameter $M$ vanishes for the
    two respective quasiclassical phases.  They represent the
    respective values $\alpha_{c_{1}}(\delta)$, $\alpha_{c_{2}}(\delta)$ and and
    $\delta_{c_{1}}^{>}(\alpha)$, $\delta_{c_{2}}^{>}(\alpha)$ [and
    also $\delta_{c_{1}}^{<}(\alpha)$, $\delta_{c_{2}}^{<}(\alpha)$
    for values of $\alpha$ in the range
    $\alpha_{c_{1}}(0) < \alpha < \alpha_{1}^{>}$ and
    $\alpha_{2}^{<}<\alpha<\alpha_{c_{2}}(0)$, respectively].  In each
    case the N\'{e}el and the striped states are used as CCM model
    states, and Eq.\ (\ref{M_extrapo_frustrated}) is used for the
    extrapolations with the respective LSUB$n$ data sets
    $n=\{2,6,10\}$ used as input.  (See also Sec.\ \ref{discuss_summary_sec} for a discussion of any possible errors in the phase diagram.)}
\label{phase_diag}
\end{figure}

Thus, finally, in Fig.\ \ref{phase_diag} we show our best estimate for
the $T=0$ phase diagram of the model in the $\alpha\delta$ plane.  For
reasons given above the phase boundaries are determined from
LSUB$\infty(1)$ points at which the magnetic order parameter $M$
vanishes.
These points are determined from extrapolating our LSUB$n$ results for
both quasiclassical AFM phases with Eq.\ (\ref{M_extrapo_frustrated}),
and using the data sets with $n=\{2,6,10\}$, which overcome any errors
associated with the $(4m-2)/4m$ staggering effects discussed above, as
input.  Points on the two (i.e., N\'{e}el and striped) phase
boundaries are shown both at fixed values of $\alpha$ and fixed values
of $\delta$.  The former [indicated by green plus $(+)$ symbols] are
obtained from curves such as those shown in Fig.\
\ref{M_J2fix-selective_extrapo}.  They thereby represent the
corresponding points $\delta_{c_{1}}^{>}(\alpha)$ [and also
$\delta_{c_{1}}^{<}(\alpha)$ for fixed $\alpha$ in the range
$\alpha_{c_{1}}(0) < \alpha < \alpha_{1}^{>}$] for the N\'{e}el phase,
and $\delta_{c_{2}}^{>}(\alpha)$ [and also
$\delta_{c_{2}}^{<}(\alpha)$ for fixed $\alpha$ in the range
$\alpha_{2}^{<} < \alpha < \alpha_{c_{2}}(0)$] for the striped phase.
The latter [indicated by red cross $(\times)$ symbols] are similarly
obtained from corresponding extrapolated curves for $M$ as a function
of $\alpha$ for various fixed values of $\delta$.  They are hence the
respective points $\alpha_{c_{1}}(\delta)$ for the N\'{e}el phase and
$\alpha_{c_{2}}(\delta)$ for the striped phase.

One can clearly see from Fig.\ \ref{phase_diag} that on both phase
boundaries the two sets of critical points, obtained from the
completely independent results at fixed values of $\alpha$ and fixed
values of $\delta$, agree extremely well with one another.  This is
definite evidence for both the consistency and high accuracy of the
extrapolation procedure that we have adopted.

Perhaps the most striking aspect of the $T=0$ phase diagram is the
marked ``avoided crossing'' behavior of the two reentrant phase
boundaries.  We note that the major portions of the upper parts of
both boundaries (i.e., for $0 \leq \alpha \lesssim 0.4$ for the
N\'{e}el case and $\alpha \gtrsim 0.6$ for the striped case) are quite
well approximated as straight lines.  Thus, if these (approximate)
straight lines were then to be extended they would cross, and the
N\'{e}el line would intersect the $\delta=0$ axis at a value rather
close to the monolayer striped QCP at $\alpha_{c_{2}}(0)$, while the
striped line would intersect the $\delta=0$ axis at a value
close to the monolayer N\'{e}el QCP at $\alpha_{c_{1}}(0)$.

\section{DISCUSSION AND SUMMARY}
\label{discuss_summary_sec}
We have studied a frustrated spin-$\frac{1}{2}$
$J_{1}$--$J_{2}$--$J_{3}$--$J_{1}^{\perp}$ Heisenberg antiferromagnet
on an $AA$-stacked honeycomb lattice in the case when $J_{1}>0$,
$J_{3}=J_{2}\equiv\alpha J_{1}>0$, and
$J_{1}^{\perp}\equiv\delta J_{1}>0$.  In particular, we have used the
CCM implemented to very high order of approximation to give an
accurate description of its $T=0$ quantum phase diagram in the window
$0 \leq \alpha \leq 1$, $0 \leq \delta \leq 1$ of the $\alpha \delta$
plane.  This window includes two quasiclassical phases with AFM
magnetic order (viz., the N\'{e}el and striped phases), plus an
intermediate paramagnetic phase that exhibits VBC order of various
types.

Within the studied window there are thus two phase boundaries,
$\delta = \delta_{c_{1}}(\alpha)$ [or, equivalently,
$\alpha = \alpha_{c_{1}}(\delta)$], along which N\'{e}el order melts,
and $\delta = \delta_{c_{2}}(\alpha)$ [or, equivalently,
$\alpha=\alpha_{c_{2}}(\delta)$], along which striped order melts.  We
have seen that these two boundaries exhibit a distinct ``avoided
crossing'' type of behavior, with both displaying a consequent
reentrant property.  In the $\alpha\delta$ window under study we found
that N\'{e}el order thus exists only for values of $\delta$ in the
range
$\delta_{c_{1}}^{<}(\alpha) < \delta < \delta_{c_{1}}^{>}(\alpha)$.
Furthermore, we found that, whereas $\delta_{c_{1}}^{<}(\alpha)=0$ in
the window for values of $\alpha$ in the range
$\alpha < \alpha_{c_{1}}(0)$, $\delta_{c_{1}}^{<}(\alpha)>0$ for
values of $\alpha$ in the respective range
$\alpha_{c_{1}}(0) < \alpha < \alpha_{1}^{>}$, where
$\delta_{c_{1}}^{<}(\alpha_{1}^{>})=\delta_{c_{1}}^{>}(\alpha_{1}^{>})$.
Similarly, we also found that in the same window striped order exists
only for values of $\delta$ in the range
$\delta_{c_{2}}^{<}(\alpha) < \delta < \delta_{c_{2}}^{>}(\alpha)$.
Comparable to the N\'{e}el phase, we also observed for the striped
phase that, whereas $\delta_{c_{2}}^{<}(\alpha)=0$ in the window under
study for values of $\alpha$ in the range
$\alpha > \alpha_{c_{2}}(0)$, $\delta_{c_{2}}^{<}(\alpha) > 0$ for
values of $\alpha$ in the corresponding range
$\alpha_{2}^{<} < \alpha < \alpha_{c_{2}}(0)$, where
$\delta_{c_{2}}^{<}(\alpha_{2}^{<})=\delta_{c_{2}}^{>}(\alpha_{2}^{<})$.
Our best estimates for the extremal points of the two quasiclassical
phases are found to be $\alpha_{1}^{>}=0.49(1)$ on the N\'{e}el phase
boundary and $\alpha_{2}^{<}=0.56(1)$ on the striped phase boundary,
where the errors are estimated from a sensitivity analysis of our
extrapolation procedure for the magnetic order parameter $M$.

Comparable errors are expected along most of the two phase boundaries
in Fig.\ \ref{phase_diag}.  The most sensitive region, however, which
is the only exception, is that close to the $\delta=0$ axis for the
N\'{e}el boundary, as we have already noted above.  For example, we have obtained the extrapolated
value $\alpha_{c_{1}}(0) \approx 0.379$ based on Eq.\
(\ref{M_extrapo_frustrated}) when used with the LSUB$n$ input data set
$n=\{2,6,10\}$.  By contrast, in an earlier CCM analysis of the
spin-$\frac{1}{2}$ $J_{1}$--$J_{2}$--$J_{3}$ honeycomb monolayer with
$J_{3}=J_{2}\equiv\alpha J_{1}$ \cite{DJJF:2011_honeycomb},
corresponding extrapolated values were obtained of
$\alpha_{c_{1}}(0)\approx 0.466$ when based on the LSUB$n$ input data set
$n=\{6,8,10,12\}$ and $\alpha_{c_{1}}(0) \approx 0.448$ when based on the
corresponding set $n=\{6,8,10\}$.  This sensitivity is surely
associated with the fact that the QPT at $\alpha_{c_{1}}(0)$ for the
monolayer, from the N\'{e}el phase to the plaquette VBC (PVBC) phase, appears
to be of continuous, deconfined type \cite{DJJF:2011_honeycomb}.  A
sensitivity analysis yields our best estimate for this monolayer QCP
to be at $\alpha_{c_{1}}(0)=0.46(2)$.  Interestingly, this value is now
even closer to the point where one would estimate the striped boundary
curve to intersect the $\delta=0$ axis if its crossing with the
N\'{e}el boundary curve would {\it not} be avoided.

By contrast, the QPT at $\alpha_{c_{2}}(0)$ for the monolayer, from the PVBC phase to be striped phase, appears to be of first-order
type \cite{DJJF:2011_honeycomb}, and our CCM estimates for it are
accordingly much less sensitive to the extrapolation LSUB$n$ input
data set.  Compared with our own value here of
$\alpha_{c_{2}}(0) \approx 0.595$ from using the LSUB$n$ set $n=\{2,6,10\}$,
an earlier CCM analysis of the spin-$\frac{1}{2}$
$J_{1}$--$J_{2}$--$J_{3}$ honeycomb-lattice monolayer with
$J_{3}=J_{2}\equiv \alpha J_{1}$ \cite{DJJF:2011_honeycomb}, gave the
corresponding value $\alpha_{c_{2}}(0) \approx 0.601$ based on both
sets $n=\{6,8,10,12\}$ and $n=\{6,8,10\}$.  Our best overall estimate
is $\alpha_{c_{2}}(0)=0.600(5)$.

It is perhaps worthwhile to end by pointing out why we can assert with
confidence that our extrapolation procedure is indeed robust, since
this is the sole approximation in all CCM calculations.  Firstly, the
method has now been used in well over 100 different papers for a wide
variety of frustrated quantum magnets (and see, e.g., Refs.\
\cite{DJJFarnell:2014_archimedeanLatt,Bishop:2017_honeycomb_bilayer_J1J2J1perp,DJJF:2011_honeycomb,Bishop:2012_honey_circle-phase,Li:2012_honey_full,Zeng:1998_SqLatt_TrianLatt,Fa:2004_QM-coll,Bishop:1994_ccm_XXZ_SqLatt,Zeng:1995_ccm_triangLatt,Zeng:1996_SqLatt_TrianLatt,Bishop:2000_XXZ,Kruger:2000_JJprime,Fa:2001_SqLatt_s1,Darradi:2005_Shastry-Sutherland,Bi:2008_EPL_J1J1primeJ2_s1,Bi:2008_JPCM_J1xxzJ2xxz_s1,Bi:2009_SqTriangle,Bishop:2010_UJack,Bishop:2010_KagomeSq,Bishop:2011_UJack_GrtSpins,PHYLi:2012_SqTriangle_grtSpins,PHYLi:2012_honeycomb_J1neg,Li:2012_anisotropic_kagomeSq,Bishop:2012_honeyJ1-J2,RFB:2013_hcomb_SDVBC,Richter:2015_ccm_J1J2sq_spinGap,Bishop:2015_J1J2-triang_spinGap,Bishop:2015_honey_low-E-param,Bishop:2016_honey_grtSpins,Li:2016_honey_grtSpins,Li:2016_honeyJ1-J2_s1,Li:2018_crossStripe_low-E-param}
and references cited therein), where the same extrapolation schemes as
used here have been utilized, and in virtually all of which the
results obtained have been shown to be either the best, or among the
best, available.  Secondly, in all of the many cases in the literature
cited, where it has been possible to compare results obtained from
different LSUB$n$ input sets, it has been shown that the obtained
extrapolants for all physical parameters agree with one another,
typically to
$\sim 1\%$ or better.  Thirdly, the same is true here in the limited
cases where we can test it.  For example, for the limiting case
$\delta=0$ of the monolayer, where
LSUB$n$ calculations can additionally be done for the case
$n=12$, the fits to all parameters stay unchanged to the same level.
Lastly, in all the limited (i.e., unfrustrated) cases where comparison
can be made with the essentially exact results of large-scale QMC
calculations (as cited here for the case
$\alpha=0=\delta$), the extrapolated CCM values for all physical
parameters typically agree with the extrapolated QMC values (i.e.,
after finite-side scaling to the $N \rightarrow
\infty$ limit) again to $\sim 1\%$ (or better).  The real point here
is that the present honeycomb bilayer model is particularly
challenging due to the unavoidable $(4m-2)/4m$ staggering effects
discussed.  While it is therefore true that we are thereby {\it
  forced} to use only few (3 or 4) points in our fits, we can
nevertheless be confident of their robustness for the reasons cited.

In conclusion, we have found that the CCM when implemented to high
orders of LSUB$n$ approximation and the results suitably extrapolated
to the (exact) $n \rightarrow \infty$ limit, is capable of giving very
accurate descriptions of the $T=0$ quantum phase boundaries of this
frustrated $AA$-stacked honeycomb bilayer model.  In the light of this
it would clearly also be of interest to use the method to study
comparable bilayer models with the staggered Bernal $AB$ stacking.

\section*{ACKNOWLEDGMENTS}
We thank the University of Minnesota Supercomputing Institute for the
grant of supercomputing facilities, on which the work reported here
was performed.  One of us (RFB) gratefully
acknowledges the Leverhulme Trust (United Kingdom) for the award of an
Emeritus Fellowship (EM-2015-007).  

\bibliographystyle{apsrev4-1}
\bibliography{bib_general}

\end{document}